\documentclass[12pt]{article}

\usepackage[titletoc,title]{appendix}
\usepackage[margin=1in]{geometry}
\usepackage{array}
\usepackage{color}
\usepackage{amsmath,bm}
\usepackage{bbm}
\usepackage{amssymb}
\usepackage{amsthm}
\usepackage{amsfonts}
\usepackage{enumerate}
\usepackage[toc]{glossaries}
\usepackage{hyperref}
\usepackage{listings}
\usepackage{natbib}
\usepackage{url}
\usepackage{graphicx}
\usepackage{graphics}
\usepackage{subcaption}
\usepackage{fancyhdr}
\usepackage{rotating}
\usepackage{multirow}
\usepackage[usenames,dvipsnames]{xcolor}
\usepackage{sectsty}
\usepackage{titlesec}
\usepackage{booktabs}
\usepackage{datetime}
\usepackage{tikz}
\usetikzlibrary{arrows,automata,positioning,chains,fit,shapes,calc,patterns}
\usepackage{type1cm}
\usepackage{lettrine}
\usepackage{comment}
\usepackage{morefloats}
\usepackage{longtable}
\usepackage{psfrag}
\usepackage{changepage}
\usepackage{xstring}
\usepackage[ampersand]{}
\usepackage{mathtools}
\usepackage{multicol}
\usepackage{setspace}
\usepackage{censor}
\usepackage{tabularx}
\usepackage[space]{grffile}
\usepackage{blkarray, bigstrut} %
\usepackage[normalem]{ulem}
\usepackage[utf8]{inputenc}
\useunder{\uline}{\ul}{}

\settimeformat{ampmtime}
\mmddyyyydate

\def\ig{\iota\gamma}
\def\g{\gamma}
\def\i{\iota}

\usepackage{multirow,bigdelim}

\IfFileExists{/Users/jfogel/Networks/bibtex/bibtex.bib}{

}{

}

\definecolor{linkcolor}{rgb}{0,0,0.50}
\setcitestyle{authoryear, round, semicolon, aysep={},yysep={;},notesep={,}}
\hypersetup{
    bookmarks=true,                 
    unicode=false,                  
    pdftoolbar=true,                
    pdfmenubar=true,                
    pdffitwindow=true,              
    pdfstartview={FitH},            
    pdftitle={},                    
    pdfauthor={BPS team},           
    pdfnewwindow=true,              
    colorlinks=true,                
    linkcolor=black,                 
    citecolor=black,                
    filecolor=magenta,              
    urlcolor=blue,                  
    breaklinks=true
}

\IfFileExists{/Users/jfogel/Networks/bibtex/bibtex.bib}{
	}{
	}

\setcitestyle{authoryear, round, semicolon, aysep={},yysep={;},notesep={,}}

\definecolor{gold}{rgb}{0.85,.66,0}
\definecolor{blue}{rgb}{0,0,1}

\def\bs{\begin{sideways}}
\def\es{\end{sideways}}

\usepackage[draft]{fixme}
\fxsetup{layout=footnote, marginclue}

\renewcommand{\P}{\mathbb {P}}

\usepackage{footmisc}

\theoremstyle{definition}

\theoremstyle{plain}

\usepackage{mathrsfs} 

\def\ve{\varepsilon}

\makeglossaries

\title{Detailed Gender Wage Gap Decompositions: \\ Controlling for Worker Unobserved Heterogeneity Using Network Theory\footnote{Fogel Opportunity Insights, jamiefogel@g.harvard.edu.  Modenesi: University of Michigan, bmodene@umich.edu. This material is based upon work supported by the National Science Foundation Graduate Research Fellowship Program under Grant No. 1256260. Any opinions, findings, and conclusions or recommendations expressed in this material are those of the author(s) and do not necessarily reflect the views of the National Science Foundation. This research is also supported by the Alfred P. Sloan Foundation through the CenHRS project at the University of Michigan. This work is done in partnership with the Brazilian Institute of Applied Economic Research (IPEA). We thank John Bound, Abigail Jacobs, Matthew Shapiro, Mel Stephens, and Sebastian Sotelo for advice and guidance throughout this project. We also thank Charlie Brown, Zach Brown, Raj Chetty, Ying Fan, John Friedman, Florian Gunsilius, Nathan Hendren, Dhiren Patki, Rafael Pereira, Matthew Staiger, Dyanne Vaught, and Jean-Gabriel Young for helpful comments and discussions.  We also received helpful feedback from seminar participants at the University of Michigan, Labo(u)r Day, the Urban Economics Association, Networks 2021, Yale University, Duke University, the Federal Reserve Bank of Boston, Opportunity Insights, and JAM.}}

\author{Jamie Fogel and Bernardo Modenesi}
\date{}

\begin{document}

\maketitle

\begin{abstract}
Recent advances in the literature of decomposition methods in economics have allowed for the identification and estimation of detailed wage gap decompositions. In this context, building reliable counterfactuals requires using tighter controls to ensure that similar workers are correctly identified by making sure that important unobserved variables such as skills are controlled for, as well as comparing only workers with similar observable characteristics. This paper contributes to the wage decomposition literature in two main ways: (i) developing an economic principled network based approach to control for unobserved worker skills heterogeneity in the presence of potential discrimination; and (ii) extending existing generic decomposition tools to accommodate for potential lack of overlapping supports in covariates between groups being compared, which is likely to be the norm in more detailed decompositions. We illustrate the methodology by decomposing the gender wage gap in Brazil. 
\end{abstract}
\clearpage


\clearpage

\onehalfspacing

\section{Introduction}

Significant attention has been paid to the gap in wages between men and women. Researchers are interested in understanding how much of the gap is due to men and women performing different work using different skills, and how much is due to men and women being paid differently for similar work.  A number of methods exist for trying to answer this question. These methods decompose gender wage gaps into a portion explained by differences in characteristics between men and women, and a portion explained by differences in the return to characteristics, or ``discrimination''. However, all of these methods rely on three assumptions. First, they assume that unobserved determinants of earnings are independent of gender. To the extent that there exist unobserved worker characteristics that are important for determining wages and are correlated with gender, then researchers will obtain  biased estimates of the return to observable characteristics. As a result, decompositions of gender wage gaps into a component explained by covariates and a component explained by the return to covariates will be incorrect. Second, they assume a functional form in order to estimate the function that maps observable characteristics into wages and thus serves as the foundation for counterfactuals that ask what men would earn if they had the same characteristics except their gender were switched to female, and vice versa. Third, they assume that the covariates for male workers and female workers share a common support. While this is likely to hold when the number of covariates is small, as more covariates are added (possibly to satisfy the independence assumption) the common support assumption becomes more likely to be violated.\footnote{As more covariates are added it becomes harder to find another worker who shares the same values of all covariates.}

In this paper, we (i) propose a new method for identifying unobserved determinants of workers' earnings from the information revealed by detailed data on worker--job matching patterns, (ii) non-parametrically estimate counterfactual wage functions for male and female workers, (iii) allow for a relaxation of the common support assumption, and (iv) apply our methods by decomposing the gender wage gap in Brazil using improved counterfactuals based on (i), (ii) and (iii). We find that the Brazilian gender wage gap is almost entirely explained by male and female workers who possess similar skills and perform similar tasks being paid different wages, not women possessing skills or tasks that pay relatively lower wages. 

To understand the problem created by unobserved determinants of productivity, suppose that there are three types of worker characteristics that are relevant for determining wages: gender, other characteristics observable to researchers, and characteristics that are observable to labor market participants, but not to researchers. A  naive wage decomposition would simply compare male wages to female wages and attribute all differences to the effect of gender. A more common approach would condition on observable characteristics like age, experience, occupation, education, and union membership and would attribute all differences in wages, conditional on these characteristics, solely to being a woman as opposed to being a man. However, this would miss the fact that even workers with identical observable covariates may perform distinct labor. As \citet{goldin2014} shows, male lawyers significantly outearn female lawyers largely because males are more likely to work long, inflexible hours, which leads to high wages. Therefore, if we simply compared the wages of male lawyers to the wages of female lawyers, we might mistakenly conclude that male and female lawyers receive differential pay for the same work, when in fact male and female lawyers perform different types of legal work. In other words, male and female lawyers differ in terms of covariates that are observed by labor market participants but not by researchers. 

The key to our approach is identifying information about worker characteristics observable to labor market participants, but not to researchers, directly from the behavior of labor market participants. If we can identify groups of workers and groups of jobs who are similar \emph{from the perspective of labor market participants}, then we can be confident that any gender wage differentials within these groups are due to differential returns to labor market activities by gender, rather than differences in the work done by male and female workers. 

We employ a revealed preference approach that relies on workers' and jobs' choices, rather than observable variables or expert judgments, to classify workers and jobs into groups. Our key insight is that linked employer-employee data contain a previously underutilized source of information: millions of worker--job matches, each of which reflects workers' and jobs' perceptions of the workers' skills and the jobs' tasks. Intuitively, if two workers are employed by the same job, they probably have similar skills, and if two jobs employ the same worker those jobs probably require workers to perform similar tasks. However, since discrimination may lead men and women with similar skills to sort into different jobs, our method includes a correction for gender-based sorting into jobs that normalizes workers' job match probabilities by the match probabilities for their gender.

We formalize this intuition and apply it to large-scale data using a \citet{Roy1951} model in which workers supply labor to jobs according to comparative advantage. Workers belong to a discrete set of latent \emph{worker types} defined by having the same ``skills'' and jobs belong to a discrete set of latent \emph{markets} defined by requiring employees to perform the same ``tasks.''\footnote{``Skills'' and ``tasks'' should be interpreted broadly as any worker and job characteristics that determine which workers match with which jobs.}  Workers match with jobs according to comparative advantage, which is determined by complementarities between skills and tasks at the worker type--market level. Workers who have similar vectors of match probabilities over markets  are therefore revealed to have similar skills and belong to the same worker type, and jobs that have similar vectors of match probabilities over worker types are revealed to have similar tasks and belong to the same market.  Our model extends the model in \citet{FogelModenesi2021} to allow firms to have labor market power, thereby rationalizing pay heterogeneity among workers with the same skills in jobs requiring the same tasks and microfounding the correction for gender-based sorting.

Once we have clustered workers with similar skills into worker types and jobs requiring similar tasks into markets, we turn to estimating counterfactual wage functions. Traditional decomposition methods estimate counterfactual female earnings by fitting wage regressions using observations for male workers only, but generating predicted values by multiplying average female covariate values by the male regression coefficients. This approach suffers from three main issues: (i) it requires the researcher to impose a restrictive regression functional form; (ii) it does not necessarily allow for heterogeneous returns to covariates in predictions; and (iii) it does not have embedded tools to handle when workers do not share similar covariate support. Taken together, these issues can potentially bias the counterfactual estimation exercise, which is the foundation of gender wage gap decompositions. In order to circumvent these issues, we make use of a flexible matching estimator for counterfactual earnings. 

We implement a matching estimator in which we match male and female workers who belong to the same worker type and are employed by jobs in the same market. In doing so, we implicitly assume that worker types and markets fully account for all factors, other than gender, that affect workers' wages, although we also estimate specifications in which we include other observable characteristics in addition to worker types and markets. Within these matched groups, we use the male workers' mean wages as counterfactuals for what the female workers would have earned if they were male, and vice versa. We compare our matching estimator to a standard estimator and find similar results, although in some specifications the matching estimator is clearly preferable. However, there may be some worker type--market cells with no male workers or no female workers so we introduce a correction to account for this lack of common support.

We address the issue of a lack of common covariate support between male and female workers by decomposing the gender wage gap into four components: (i) differences due to different covariate distributions between groups, i.e. the \textit{composition factor}, for observations that share the same support; (ii) differences related to differential returns to covariates between groups over a common support of the covariates, i.e. the \textit{structural factor}, often associated with labor market discrimination; (iii) a part due to observations from male workers being out of the female workers' support of the covariates; and (iv) the last portion related to observations of female workers being out of the male workers' support of the covariates. This decomposition allows us to perform counterfactuals similar to existing methods for the part of the distribution of the covariates for which male and female workers have common support, yet it still allows us to quantify how much of the gender wage gap occurs outside the region of common support and would therefore be ignored by standard decomposition methods. 

We estimate our model and conduct empirical analyses using Brazilian administrative records from the Annual Social Information Survey (RAIS) that is managed by the Brazilian labor ministry. The RAIS data contain detailed information about every formal sector employment contract, including worker demographic information, occupation, sector, and earnings. Critically, these data represent a network of worker--job matches in which workers are connected to every job they have ever held, allowing us to identify job histories of workers, their coworkers, their coworkers' coworkers, and so on. We restrict our analysis to the Rio de Janeiro metropolitan area both for computational reasons and because restricting to a single metropolitan area enables us to focus on skills and tasks dimensions of worker and job heterogeneity rather than geographic heterogeneity.

In our data, the average male worker earns a wage 16.7\% higher than the average female worker. Our primary result is that almost the entire gender wage gap is attributable to male and female workers who possess similar skills and perform similar tasks being paid differently, or what is often referred to as ``discrimination.'' This is true at the aggregate level, and remains true when we perform wage decompositions within each worker type--market cell, indicating that this is a widespread phenomenon, not one driven by large wage differentials in small subsets of the labor market. We find that wage decompositions based on standard observable variables suffer from omitted variable bias, emphasizing the need for detailed worker and job characteristics in the form of worker types and markets. We find that wage decompositions based on linear regressions yield similar findings to those based on matching when a lack of common support is not an issue, however when male and female workers' characteristics do not share a common support the matching estimator with corrections for a lack of common support outperforms alternatives.

\textbf{Literature:} The literature of decomposition methods in economics can be classified into two main branches. The first decomposes average differences in a variable of interest $Y$ --- often wages --- between two groups of workers. The most widespread method in this class was developed by \citet{oaxaca1973} and \citet{blinder1973}. The second branch decomposes functionals of the variable of interest $Y$ -- e.g. its distribution or quantile function. Given that functionals of a variable often provide more information than its average, the second group of decompositions is referred to as ``detailed decompositions'' (\citealp{fortinlemieuxfirpo2011}). A seminal paper in this group is \citet{dinardofortinlemieux1996}\footnote{\citet{barskyboundcharleslupton2002} develop a methodology similar to \citet{dinardofortinlemieux1996}, focusing on issues that arise from lack of common covariate support between the groups in the decomposition. \citet{Modenesi2022} discusses their approach in light of alternatives to handle the lack of common support.} and their methodology and inference was further generalized and improved later by \citet{chernozhukovfernandezvalmelly2013}\footnote{\citet{firpofortinlemieux2018} later in this literature uses influence functions to propose a detailed decomposition that is invariant to the order of the decomposition.}. We follow the first branch of the literature in focusing on average differences, largely because our rich set of controls introduces a curse of dimensionality that renders detailed decompositions infeasible.

Our method for handling a lack of common covariate support follows \citet{nopo2008} and \citet{garcianoposalardi2009}\footnote{\citet{garcianoposalardi2009} and \citet{morelloanjolim2021} both study the evolution of the Brazilian gender gap. \citet{garcianoposalardi2009} uses the same approach we use to handle the problem of lack of overlapping supports, and \citet{morelloanjolim2021} have a similar matching methodology to decompose the gender gap. In addition to using similar methods for the decomposition, we add the skills and tasks controls derived from the labor market network, and we derive a distribution of gender gaps for different clusters of similar workers performing similar tasks.}. In concurrent work we extend \citet{nopo2008} to generic ``detailed decompositions'' \citep{Modenesi2022}.

Our model of labor market power builds on \citet{cardcardosokline2015}, \citet{cardcardosoheiningkline2018} and \citet{gerardlagosseverninicard2018} but allows for significantly more granular worker and job heterogeneity. The way we model multidimensional worker--job heterogeneity relates to papers that use a skills-tasks framework in the worker-job matching literature \citep{AutorLevyMurnane2003,AcemogluAutor2011,Autor2013,Lindenlaub2017,Tan2018,Kantenga2018}. Our method for clustering workers and jobs fits into the relatively recent literature in labor economics that extracts latent information from the network structural of the labor market \citep{Sorkin2018,Nimczik2018,JaroschNimczikSorkin2019} and directly extends \citet{FogelModenesi2021} by allowing for labor market power. Methodologically, we draw from the community detection branch of network theory \citep{LarremoreClausetJacobs2014,peixoto2018,Peixoto2019}\footnote{More precisely, we employ a variant of the SBM which makes use of network edge weights (\citealp{peixoto2018}), which are key for us to model the presence of potential discrimination in the labor market.}. Our paper connects to this literature by formalizing a theoretical link between monopsonistic labor market models and the stochastic block model, providing microfoundations and economic interpretability of network theory unsupervised learning tools in order to solve economic problems.

By controlling for skills and tasks, our papers share common ground with \citet{goldin2014} and \citet{hurstrubinsteinshimizu2021}. \citet{goldin2014} indicates that the potential residual discrimination in the gender wage gap is due to the nature of the tasks in some occupations, by using a linear regression approach dummies for occupation interacted with the gender dummy. We add to her approach by proposing an economic model for discrimination, which provides us with \textit{both} worker and job heterogeneity controls, in addition to performing the gender gap decomposition while taking into account potential violations of conventional decomposition assumptions. \citet{hurstrubinsteinshimizu2021} on the other hand are assessing the black-white wage gap over time as function of changes in the taste vs statistical discrimination factors, as well as the result of workers sorting after these changes.

\textbf{Roadmap:} The paper proceeds as follows. Section \ref{sec:decomposition_motivation} introduces a simple framework for decomposition methods. Section \ref{sec:model_and_sbm} presents our model of worker--job matching and derives from it our algorithm for clustering workers into worker types and jobs into markets. Section \ref{sec:wage_gap_decomposition} provides greater detail on the wage gap decomposition methods we employ. Section \ref{sec:data_ch2} describes our data. Section \ref{sec:results} presents results. Finally, Section \ref{sec:conclusion_ch2} concludes.


\section{A framework for decomposition methods}\label{sec:decomposition_motivation}

We introduce a simple framework for decomposition methods to guide the analysis in this paper. Define the actual wage of worker $i$ employed by job $j$ as $Y_{ij}$, and let $G_i$ be a dummy denoting whether worker $i$ is male. The difference between the average wage for male workers and the average wage for female workers, which we call the ``overall wage gap,'' can be expressed as:

\begin{equation}\label{eq:overall_gap}
    \Delta := E[Y_{ij} | G_i = 1] - E[Y_{ij} | G_i = 0]
\end{equation}

The overall wage gap above can be decomposed into two factors: differences in productivity between male and female workers, usually referred to as the \textit{composition} factor; and differences in pay between equally productive male and female workers, known as the \textit{structural} factor. We use the potential outcomes framework in order to formally decompose the overall wage gap into these two factors. Denote by $Y_{0ij}$ the potential wage of worker $i$ employed by job $j$ when the worker is female, and $Y_{1ij}$ the potential wage of worker $i$ employed by job $j$ when the worker is male.  Let $x$ be the vector of all variables that determine workers' productivity. We assume that the worker's gender may affect their pay, but does not directly affect their productivity. We  represent the potential outcomes as functions of $x$ as follows: $Y_{gij} := Y_{g}(x_{ij}), g \in \{0,1\}$. Notice that $x$ has both $i$ and $j$ subscripts, as the marginal product of worker $i$ at their current job $j$ depends on both the worker's skills and the job's tasks. The fact that there is a different earnings function for men and women reflects the possibility that male and female workers with identical productivities may be paid differently. Furthermore, it is possible to use the dummy for gender to represent observed wages as a function of potential outcomes using a switching regression model $Y_{ij} := G_i Y_{g}(x_{ij}) - (1 - G_i) Y_{g}(x_{ij})$.

At this point we are able to decompose the overall wage gap, $\Delta$, into the \textit{composition} and \textit{structural} components mentioned above by adding and subtracting the quantity\footnote{Analogously, the overall decomposition can be performed by adding and subtracting the male counterfactual quantity $E[Y_{0}(x_{ij}) | G_i = 1]$ to $\Delta$. The main results in this paper use the female counterfactual approach.} \[E[Y_{1}(x_{ij}) | G_i = 0] := \int Y_{1}(x_{ij}) dF_{G=0}(x)\] 
from the overall wage gap $\Delta$, where $F_{G=0}(x)$ is the productivity distribution for female workers.  Intuitively, $E[Y_{1}(x_{ij}) | G_i = 0]$ is the mean earnings for a counterfactual set of workers possessing the female productivity distribution, but who are paid like men\footnote{Alternatively, this counterfactual term can be interpreted as the mean earnings of male workers whose productivity distribution was adjusted to match the female productivity distribution}. 

\begin{equation}\label{eq:overall_gap_decomp}
    \Delta := \underset{\Delta_X := \text{Composition}}{\underbrace{E[Y_{ij} | G_i = 1] - E[Y_{1}(x_{ij}) | G_i = 0]}} + \underset{\Delta_0 := \text{Structural}}{\underbrace{E[Y_{1}(x_{ij}) | G_i = 0] - E[Y_{ij} | G_i = 0]}} 
\end{equation}

The \textit{composition} portion can be rewritten as $E[Y_1(x_{ij}) | G_i = 1] - E[Y_1(x_{ij}) | G_i = 0]$\footnote{We use the representation of the observed $Y$ in terms of potential outcomes to write $E[Y_{ij} | G_i = 1] = E[G_i Y_{g}(x_{ij}) - (1 - G_i) Y_{g}(x_{ij})| G_i = 1] = E[Y_1(x_{ij}) | G_i = 1]$ and substitute it in $\Delta_X$.}. It represents the difference between what male workers actually earn and what male workers would have earned in a counterfactual scenario in which their productivity distribution was equivalent to the female productivity distribution. This quantity captures the portion of the overall wage gap attributable to differences in the composition, or distribution of productivity, between male and female workers. The \textit{structural} portion is equivalent to $E[Y_1(x_{ij}) - Y_0(x_{ij}) | G_i = 0]$\footnote{Analogously to the previous term, using the map from the potential outcomes to the observed $Y$, we can write  $E[Y_{ij} | G_i = 0] = E[G_i Y_{g}(x_{ij}) - (1 - G_i) Y_{g}(x_{ij})| G_i = 1] = E[Y_0(x_{ij}) | G_i = 0]$ and substitute it in $\Delta_0$.}. This is the difference between female earnings in a counterfactual state in which females were paid equivalently to what equally productive male workers are paid and actual average female earnings. This portion of the overall wage gap is due to structural differences in how the two genders are paid, holding productivity constant, which is why this term is often associated with a form of discrimination. 

What we define as the structural component might reasonably be thought as discrimination, where labor market discrimination is defined as workers with similar productivity, performing similar tasks, and being paid differently based on observables that do not influence productivity. Other forms of discrimination may exist --- including mistreatment or harassment, differential pre-job human capital accumulation opportunities, or discriminatory hiring practices --- but we do not consider those in this paper. In our set up, individual discrimination occurs when the wage for worker $i$ at job $j$ is different if the individual's gender changes, \textit{ceteris paribus}, i.e. $Y_{1}(x_{ij}) - Y_{0}(x_{ij}) \neq 0$. The problem is that, in order to measure this quantity, we run into the fundamental problem of causal inference: it is impossible to observe the potential wages in both states for the same individual. Therefore we must make assumptions in order to construct counterfactual values, i.e. the value of $Y_1$ for a female worker, or the value of $Y_0$ for a male worker. In this paper, we break the assumptions needed for the counterfactual estimation into two parts and we show how our approach contributes to deal with limitations in each of them.

The first assumption is that workers with the same values of $x$ are equally productive and would be paid equal wages if gender played no role in wage determination, conditional on productivity. This is equivalent to assuming that $x$ contains all factors that affect productivity and are correlated with gender.  This ``conditional independence/ignorability'' assumption, is the basis of all decomposition methods in economics (\citealp{fortinlemieuxfirpo2011}), as it is a requirement for consistency of its estimates for the gap decomposition portions. However, not all factors that theoretically should be included in $x$ are observable.

A problem would arise if certain factors that contribute to worker $i$'s productivity in job $j$ are both unobserved by the econometrician and correlated with gender. If such factors exist, our counterfactuals would be invalid. Specifically, wage differentials due to unobserved differences in skills and tasks between male and female workers would be attributed to the effect of gender itself. For example, if women tend to have better social skills but we do not observe social skills, then we would interpret women outearning men in social skill-intensive jobs as discrimination against men, when in fact it is simply the result of differences in unobserved skills. Therefore, it is critical to come as close as possible to identifying groups of male and female workers who have exactly the same skills and perform exactly the same tasks. If we do so, then any gender wage differentials within this group are attributable to the effect of gender \textit{per se}. In Section \ref{sec:model_and_sbm} we address this issue by identifying latent worker and job characteristics relevant to productivity and wage determination using the network of worker--job matches. 

The second set of assumptions required to build the counterfactual $Y_1(x)$ for females in $\Delta$ are related to the choice of an estimation strategy for the function $Y_1(\cdot)$\footnote{Another approach decomposes the wage \textit{distributions}, as opposed to actual wages, which would be equivalent to switching $Y$ for its distribution $F_Y$, but still needing the estimation of the counterfactual $F_{Y_1}(y|x)$ for females (e.g. \citealp{dinardofortinlemieux1996} and \citealp{chernozhukovfernandezvalmelly2013}). We choose not to employ these decompositions in this paper as our setup does not satisfy basic conditions for decomposing distributions, such as having a low-dimensional vector of observable characteristics $x$ -- given curse of dimensionality -- and having the overlapping supports assumptions satisfied.}. A common estimation strategy requires fitting a linear wage regression for males and using its estimated coefficients to predict wages, but inputting female workers' covariates (\citealp{oaxaca1973} and \citealp{blinder1973}). This approach is highly tractable, however the assumption of a linear functional form is to some extent arbitrary, and using the same regression coefficients to predict counterfactual earnings for distinct female workers (i.e. allowing no heterogeneous returns to observable characteristics) could lead to biased estimates of counterfactual earnings. An alternative approach relies on matching males to each female worker based on similar observable characteristics, and uses the wages of matched male workers in order to inform each female's counterfactual wage. This less-parametric approach has the advantage of not imposing any functional form assumption for $Y_1(\cdot)$, however it requires us to observe a sufficiently rich set of observable variables that male and female workers with the same observables may be assumed to have similar productivity. Moreover, matching methods are unreliable when we are unable to find a female worker with the same observables as a male worker, or vice versa. In Section \ref{sec:model_and_sbm} we describe a new method to enhance the set of observable characteristics available to the researcher, reducing the scope for unobserved determinants of productivity to cause biased estimates. In Section \ref{sec:wage_gap_decomposition} we compare and contrast different methods to decompose the gender wage gap given a set of observable characteristics, circumventing issues present in counterfactual earnings estimation.


\section{Revealing latent worker and job heterogeneity using network theory}\label{sec:model_and_sbm}

In this section we present an economic model of monopsonistic wage setting, which rationalizes a wage gap between two groups of workers who have different demographic characteristics, but have the same skills and perform the same tasks. Intuitively, otherwise identical male and female workers may supply labor to individual jobs with different elasticities, and jobs respond by offering them wages with different markdowns.  If one group of workers supplies labor to jobs more inelastically, then they will be paid less, holding productivity constant. Moreover, the model microfounds our network-based clustering algorithm, which identifies groups of male and female workers with similar skills who perform similar tasks, and therefore can serve as good counterfactuals for each other.  The model builds on the model of the labor market developed in \citet{FogelModenesi2021}, with two important differences: (i) in this paper workers have idiosyncratic preferences over individual jobs, not just markets, causing jobs to face upward-sloping labor supply curves, and (ii) firms may offer different wages to men and women, even if they have identical skills and perform identical tasks. The model defines a probability distribution that governs how workers match with jobs, forming the network of worker-job matches observed in linked employer-employee data. We use this probability distribution to assign similar workers to worker types and similar jobs to markets, using a Bayesian method based on generative network theory models, which we present after the economic model.

\subsection{Economic model}

We propose a model with two primary components: heterogeneous workers who supply labor and firms that produce goods by employing labor to perform tasks. Workers supply their skills to jobs, which are bundles of tasks embedded within firms. Jobs' tasks are combined by the firms' production functions to produce output. We assume that firms face an exogenously-determined demand for their goods\footnote{For an alternative version of the model with endogenous product demand, see \citet{FogelModenesi2021}.}. Our model of the labor market has the following components:
\begin{itemize}
    \item Each worker is endowed with a ``worker type,'' and all workers of the same type have the same skills.
    \item A job is a bundle of tasks within a firm. As we discuss in Section \ref{sec:data_ch2}, we define a job in our data as an occupation--establishment pair.
    \item Each job belongs to a ``market,'' and all jobs in the same market are composed of the same bundle of tasks. 
    \item There are $I$ worker types, indexed by $\i$, and $\Gamma$ markets, indexed by $\g$. 
    \item The key parameter governing worker-job match propensity is an $I\times \Gamma$ productivity matrix, $\Psi$, where the ($\i,\g$) cell, $\psi_{\ig}$ denotes the number of efficiency units of labor a type $\i$ worker can supply to a job in market $\g$.\footnote{We can think of $\psi_{\ig}$ as $\psi_{\ig}=f(X_{\i},Y_{\g})$, where $X_{\i}$ is an arbitrarily high dimensional vector of skills for type $\i$ workers, $Y_{\g}$ is an arbitrarily high dimensional vector of tasks for jobs in market $\g$, and $f()$ is a function mapping skills and tasks into productivity. This framework is consistent with \citet{AcemogluAutor2011}'s skill and task-based model, and is equivalent to \citet{Lindenlaub2017} and \citet{Tan2018}. A key difference is that \citeauthor{Lindenlaub2017} and \citeauthor{Tan2018} observe $X$ and $Y$ directly and assume a functional form for $f()$, whereas we assume that $X$, $Y$, and $f()$ exist but are latent. We do not identify $X$, $Y$, and $f()$ directly because in our framework $\psi_{\ig}$ is a sufficient statistic for all of them. } 
\end{itemize}
Time is discrete, with time periods indexed by $t \in \{1,\dots,T\}$ and workers make idiosyncratic moves between jobs over time. Neither workers, households, nor firms make dynamic decisions, meaning that the model may be considered one period at a time.  We do not consider capital as an input to production. 

\subsubsection{Firm's problem}

Each firm, indexed by $f$, has a production function $Y_f(\cdot)$ which aggregates tasks from different labor markets, indexed by $\g$. Firm $f$ faces exogenously-determined demand for its output, $\bar Y_f$. The firm's only cost is labor. As we discuss in the next subsection, firms face upward-sloping labor supply curves and therefore have wage-setting power. Firms demand labor in each market, $\g\in\{1,\dots,\Gamma\}$ and offer a different wage per efficiency unit of labor for each market. Firms also may offer different wages to workers in different demographic groups $g\in\{A,B\}$ (e.g. male and female workers), although type $A$ and type $B$ workers belonging to the same worker type $\i$ are equally productive in all jobs. We define a job $j$ as a firm $f$ -- market $\g$ pair. We define the wage per efficiency unit of labor for demographic group $g$ workers employed in job $j$ $w_j^g$. Define $L_j^g$ as the quantity of efficiency units of labor supplied by demographic group $g$ workers to job $j$. 

The firm's problem is to choose the quantity of labor inputs in each job for each demographic group in order to minimize costs subject to the constraint that production is greater than or equal to the firm's exogenous product demand, $\bar Y_f$: 
\begin{flalign*}
    \min_{ \{ {w}_{j}^A, {w}_{j}^B\}_{j = 1}^\Gamma }\sum_{j=1}^\Gamma w^A_{j} L^A_{j} + w^B_{j} L^B_{j}  \quad \text{s.t.} \quad Y_f \left(L_{1}, \ldots, L_{\Gamma} \right) \geq \bar Y_f
\end{flalign*}
where $L_{j} = L^A_{j} +L^B_{j}$ is the total amount of efficiency units of labor employed by job $j$ and $Y_f$ is a concave and differentiable production function.

Taking the first order condition with respect to $w_{j}^g$ allows us to solve for the wage paid by job $j$ to workers in demographic group $g$ as a markdown relative to the marginal revenue product of labor: 
\begin{flalign}
    w_{j}^g =& \underset{\text{  Markdown   }}{\underbrace{\frac{e_{j}^g}{1 + e_{j}^g}}} \qquad \times \underset{\text{Marg. revenue  product of labor}}{\underbrace{\mu_f \frac{\partial Y_f}{\partial L_{j}}}} \label{eq:markdown}
\end{flalign}
where $\mu_f$ is the shadow revenue associated with one more unit of output and $e_{j}^g := \frac{\partial L_{j}^g}{\partial w_{j}^g} \frac{w_{j}^g}{L_{j}^g} $ is the labor supply elasticity of workers from group $g$ to job $j$. 

Equation (\ref{eq:markdown}) shows that the wage paid to demographic group $g$ workers employed in job $j$ (equivalently, employed in market $\g$ by firm $f$) is the product of a markdown and the marginal revenue product of labor in job $j$. The markdown depends on the demographic group $g$'s elasticity of labor supply to job $j$. As labor supply becomes more elastic, the markdown converges to 1 and the wage converges to the marginal product of labor. Conversely, as labor supply becomes less elastic, the wage declines further below the marginal product of labor. This equation rationalizes different demographic groups being paid different wages for the same labor: if one demographic group supplies labor more inelastically, they will be paid less.\footnote{We are referring to the elasticity of labor supply \emph{to a specific job $j$}, which may differ from a group's labor supply elasticity to the overall labor market. For example, it could be the case that men supply labor more inelastically at the extensive margin, but women have stronger idiosyncratic preferences for specific jobs, making them less likely to change jobs in response to a wage differential. In this case, women would supply labor less elastically to a specific job $j$ and thus receive lower wages.}  The firm employs workers in both demographic groups despite paying them different wages because in order to attract the marginal worker from the lower-paid demographic group, it must raise wages for all inframarginal workers in that group. At some point the marginal cost (inclusive of the required raises for inframarginal workers) of hiring workers from the lower-paid demographic group exceeds the marginal cost of hiring workers from the higher-paid demographic group, and the firm will switch to hiring the higher-paid workers.  

\subsubsection{Worker's problem}
A worker belonging to worker type $\i$ and demographic group $g \in \{A,B\}$, has a two step decision. First, she chooses a market $\g$ in which to look for a job, and second she chooses a firm $f$ (and by extension a job $j$). The worker's type defines their skills. Type $\i$ workers can supply $\psi_{\ig}$ efficiency units of labor to any job in market $\g$. $\psi_{\ig}$ is a reduced form representation of the skill level of a type $\i$ worker in the various tasks required by a job in market $\g$. Units of human capital are perfectly substitutable, meaning that if type 1 workers are twice as productive as type 2 workers in a particular market $\g$ (i.e. $\psi_{1\g} = 2\psi_{2\g}$), firms would be indifferent between hiring one type 1 worker and two type 2 workers at a given wage per efficiency unit of labor, $w_{j}$. Therefore, the law of one price holds within each demographic group for each job, and a type $\i$ worker belonging to demographic group $g$ employed in a job in market $\g$ is paid $\psi_{\ig}w_j^g$. Because workers' time is indivisible, each worker may supply labor to only one job in each period and we do not consider the hours margin.

Workers choose job $j$, equivalent to $\g f$, in order to maximize utility, which is the sum of log earnings $ \log (\psi_{\ig} w_j^g) $  and an idiosyncratic preference for job $j$, $\ve_{i j}^g$:
\begin{flalign*}
    j^{*} 		=& \arg\max_{j} \quad \log (\psi_{\ig} w_j^g) + \ve_{i j}^g.  
\end{flalign*}
We assume that $\ve_{i j}^g$ follows a nested logit distribution with parameter $\nu_{\g}^g$, with the $\g$ subscript indicating that nests are defined by $\g$:
\begin{flalign*}
    \ve_{i j}^g \sim NestedLogit(\nu_{\g}^g)
\end{flalign*}   
It follows from this assumption about the distribution of $\ve_{ij}^g$ that the probability that worker $i$ belonging to worker type $\i$ and demographic group $g$ matches with job $j$ in market $\g$ is\footnote{Details for the derivation of the choice probability in the Appendix \ref{app:ch2:choice-prob}.}:
\begin{flalign}\label{eq_worker_choice}
    P(j = j^* | j \in \g, i \in \i, g) 			&= \underset{\underset{\text{\textbf{1st step}: market choice}}{\underbrace{\scriptstyle P(\g = \g^* | i \in \i, j\in\g,g)}}}{\underbrace{\frac{\exp(I_{\ig}^g)^{\nu_{\g}^g}}{\sum_{\g} \exp(I_{\ig}^g)^{\nu_{\g}^g}}}}  \underset{\underset{\text{\textbf{2nd step}: job choice}}{\underbrace{\scriptstyle P(j = j^* | i \in \i, j \in \g,\g = \g^*, g)}}}{\underbrace{	\frac{(\psi_{\ig} w_j^g)^\frac{1}{\nu_{\g}^g}}{\sum_{j \in \g} (\psi_{\ig} w_j^g)^\frac{1}{\nu_{\g}^g}}}}
\end{flalign}
where $I^g_{\ig} := \sum_{j \in \g} (\psi_{\ig} w_j^g)^\frac{1}{\nu_\g^g}$, also referred to as the inclusive value, is the expected utility a type $\i$ worker faces when choosing market $\g$. Intuitively, the nested logit assumption decomposes the job choice probability into a first stage in which the worker chooses a market and then a second stage in which the worker chooses a job conditional on their choice of a market.

\subsection{Identifying worker types and markets}

\subsubsection{Deriving the likelihood}

Now that we have derived the probability of worker $i$ matching with job $j$ from the primitives of our model, the next step is using this probability as the basis for a maximum likelihood procedure that assigns workers to worker types and jobs to markets based on the observed set of worker--job matches. This procedure builds on \citet{FogelModenesi2021}, by allowing workers in the same worker type but different demographic groups to have different vectors of match probabilities over jobs. 

We decompose the choice probability in equation (\ref{eq_worker_choice}) into a component that depends only on variation at the $\i,\g,g$ level and a component that depends on wages at individual jobs:
\begin{flalign}\label{eq_worker_choice_separation}
    P(j = j^* | j \in \g, i \in \i, g) 			&= \underset{\underset{\i-\g-g \text{ component}\quad}{\underbrace{=: \Omega_{\ig}^g}}}{\underbrace{\frac{\exp(I_{\ig}^g)^{\nu_{\g}^g-1}}{\sum_{\g} \exp(I_{\ig}^g)^{\nu_{\g}^g}}  \psi_{\ig}^\frac{1}{\nu_{\g}^g} }}   \underset{\underset{\quad j-g \text{ component}}{\underbrace{=: d_j^g  }}}{\underbrace{\vphantom{\frac{\exp(I_{\ig}^g)^{\nu_{\g}^g-1}}{\sum_{\g} \exp(I_{\ig}^g)^{\nu_{\g}^g}}  \psi_{\ig}^\frac{1}{\nu_{\g}^g}} (w_j^g)^\frac{1}{\nu_{\g}^g} }}.
\end{flalign}
The first term reflects workers choosing markets according to comparative advantage, while the second captures the fact that some jobs in market $\g$ require more workers than others (due to exogenous product demand differences), and since jobs face upward-sloping labor supply curves, they must pay higher wages to attract greater numbers of workers. Isolating the group-level ($\i,\g,g$) variation from the idiosyncratic job-level variation allows us to cluster workers into worker types and jobs into markets on the basis of having the same group-level match probabilities, as we discuss below.  

The choice probabilities we have discussed thus far refer to a single job search for worker $i$. In reality, we may observe workers searching for jobs multiple times, and each of these searches is informative about the latent worker skills and job tasks that define worker types $\i$ and markets $\g$. We incorporate repeated searches by assuming that workers periodically receive exogenous separation shocks which arrive following a Poisson process. Upon receiving a separation shock, the worker draws a new $\ve_{ij}^g$ shock and repeats the job choice process described above. Assuming that $Poisson$-distributed exogenous separations happen at a rate $d_i^g$ for the individual worker $i$, then the expected number of times she will match with job $j$ throughout our sample period is given by
\begin{flalign}\label{eq_avg_worker_job_matches}
    d_i^g \cdot P(j = j^* | j \in \g, i \in \i, g) = \Omega_{\ig}^g d_i^g d_j^g.
\end{flalign}


Equation \ref{eq_avg_worker_job_matches} forms the basis of our algorithm for clustering workers into worker types and jobs into markets, but before proceeding we must define some notation. Let $N_W$ and $N_J$ denote the number of workers and jobs, respectively, in our data. Define $A_{ij}$ as the number of times that worker $i$ is observed to match with job $j$. Further, define $\bm{A}$ as the matrix with typical element $A_{ij}$. $\bm{A}$ is a $N_W\times N_J$ matrix and represents the full set of worker--job matches observed in our data. As discussed previously, each individual worker belongs to a latent worker type denoted by $\i$ and each job belongs to a latent market denoted by $\g$. The list of all latent worker type and market assignments is stored in the $(N_W + N_J) \times 1$ vector denoted by $\bm{b}$, known as the \textit{node membership} vector. We define $\bm{g}$ as the $N_W\times1$ vector containing each worker's demographic group affiliation. The matrix of worker--job matches $\bm{A}$ and workers' demographic groups $\bm{g}$ are the data we use to cluster workers and jobs, while the node membership vector $\bm{b}$ is the latent object identified by the maximum likelihood procedure we discuss below.

Following equation (\ref{eq_avg_worker_job_matches}), the expected number of matches between a worker--job pair, $A_{ij}$, can be written as\footnote{It is worth mentioning that: (i) the information $i \in \iota, j \in \gamma$ is contained in $\bm{b}$; and (ii) $A_{ij}$ is the number of matches between worker $i$ and job $j$, which makes the event that $j=j^*|i$ equivalent to the event that $A_{ij}=1$. These two facts allow us to use more succinct notation that directly links theoretical objects in our model to data: $P(j = j^* | j \in \g, i \in \i, g) = P(A_{ij} = 1 | \bm{b}, g)$, which we know the distributional form for. This connects notations from the economic model to the network model, but it still lacks the precise definition of the likelihood of interest, $P(\bm{A}, \bm{g} | \bm{b})$, where $A_{ij}$ can assume values other than just $1$.} 
\begin{flalign} \label{eq:E_Aij}
    E[A_{ij} | \bm{b}, g] = \Omega_{\ig}^g d_i^g d_j^g.
\end{flalign}
We prove in Appendix \ref{app:poisson_proof} that our assumption of Poisson-distributed exogenous separation shocks implies that $A_{ij}$ follows a Poisson distribution:
\begin{flalign}\label{eq_choice_A}
    A_{ij} | \bm{b}, g \sim Poisson(\Omega_{\ig}^g d_i^g d_j^g) 
\end{flalign}
Finally, we incorporate equation (\ref{eq_choice_A}) above to fully characterize the likelihood of our data as a function of the unknown parameters, by applying Bayes rule: 
\begin{flalign}\label{eq_original_mle}
    P(A_{ij}, g | \bm{b}) = \underset{Poisson(\Omega_{\ig}^g d_i^g d_j^g)}{\underbrace{P(A_{ij} | \bm{b}, g)}} \underset{\alpha_{\ig}^g}{\underbrace{P(g | \bm{b})}},
\end{flalign}
where $\alpha_{\ig}^g\equiv P(g | \bm{b})$ is the fraction of type $\i$ workers employed in market $\g$ jobs who belong to the demographic group $g$. Equation \ref{eq_original_mle} corresponds to a commonly-used method from network theory known as the bipartite degree-corrected stochastic block model with edge weights (SBM). The SBM clusters \textit{nodes} in a network (workers and jobs) into groups (worker types and markets) based on patterns of connections between nodes.\footnote{\citet{LarremoreClausetJacobs2014} lays out the advantages of using bipartite models over using one-sided network projections to fit SBMs; \citet{KarrerNewman2011} presents the methodology for degree-correction as it enhances significantly the ability of the SBM to fit large scale real world networks; and \citet{peixoto2018} deal with weighted SBM inference, which is how we accommodate discrimination influencing matches within the SBM.}. The main parameter of interest is the set of assignments of workers to worker types and jobs to markets contained in $\bm{b}$, while all of the other parameters are nuisance parameters that can be straightforwardly determined after $\bm{b}$ is defined (\citealp{KarrerNewman2011}). The next step is to maximize the likelihood defined in equation \ref{eq_original_mle}, which we address in the next subsection.

\subsubsection{A Bayesian approach to recovering worker types and markets}
In order to make the estimation of worker types and markets feasible, together with using a principled method for choosing the number of clusters, we employ Bayesian methods from the network literature (\citealp{Peixoto2017}). We can rewrite equation (\ref{eq_original_mle}) as
\begin{flalign}\label{eq_posterior}
    P(\bm{b} | A_{ij}, g)  \quad	\propto& \qquad 	P(A_{ij}, g | \bm{b}) P(\bm{b}) \nonumber \\
    =& \quad 			\underset{Poisson(\Omega_{\ig}^g d_i^g d_j^g)}{\underbrace{P(A_{ij} | \bm{b}, g)}} \underset{\alpha_{\ig}^g}{\underbrace{P(g | \bm{b})}} \underset{\text{Prior}}{\underbrace{P(\bm{b})}} 
\end{flalign}
Maximizing the posterior distribution means assigning individual workers to worker types $\i$ and jobs to markets $\g$. The basic intuition follows from and is described in greater detail in \citet{FogelModenesi2021}: workers belong to the same worker type if they have approximately the same vector of match probabilities over jobs, while jobs belong to the same market if they have approximately the same vector of match probabilities over workers. The key difference in this paper is that workers in the same worker type $\i$ may belong to different demographic groups $g$ and each worker type--demographic group pair may face its own wage and therefore have its own match probability. Equation (\ref{eq_posterior}) allows for this by allowing the match probabilities $P(A_{ij}, g | \bm{b}) $ to depend on the workers' demographic group $g$ in addition to the worker types and markets stored in $\bm{b}$. 

If worker types are defined by having common vectors of match probabilities over jobs, but match probabilities are allowed to vary by demographic group within a worker type, how do we know that type $\i$ workers in group $A$ belong to the same worker type as type $\i$ workers in group $B$? The answer is embedded in equation (\ref{eq_posterior}). The $\alpha_{\ig}^g$ term in equation (\ref{eq_posterior}) adjusts workers' match probabilities so that they are relative to their own gender. Suppose women are significantly underrepresented in construction jobs and overrepresented in nursing jobs, and vice versa for men. Once we incorporate this adjustment, we would assign workers to a construction-intensive worker type if they are disproportionately likely to match with construction jobs, \textit{relative to other workers of their gender}. Once we adjust the raw match probabilities to account for this selection, we obtain identical \textit{adjusted}  match probability vectors for this group of men and this group of women, causing us to assign them to the same worker type, $\i$.

Equation (\ref{eq_posterior}) assumes that we know the number of worker types and markets \emph{a priori}, however this is rarely the case in real world applications. Therefore we must choose the number of worker types and markets, $I$ and $\Gamma$ respectively. We do so using the principle of minimum description length (MDL), an information theoretic approach that is commonly used in the network theory literature. MDL chooses the number of worker types and markets to minimize the total amount of information necessary to describe the data, where the total includes both the complexity of the model conditional on the parameters \emph{and} the complexity of the parameter space itself. MDL will penalize a model that fits the data very well but overfits by using a large number of parameters (corresponding to a large number of worker types and markets), and therefore requires a large amount of information to encode it. MDL effectively adds a penalty term in our objective function, such that our algorithm finds a parsimonious model. See \citet{FogelModenesi2021} for greater detail. 

Equation (\ref{eq_posterior}) defines a combinatorial optimization problem. If we had infinite computing resources, we would test all possible  assignments of workers to worker types and jobs to markets and choose the one that maximizes the likelihood in equation (\ref{eq_posterior}), however this is not computationally feasible for large networks like ours. Therefore, we use a Markov chain Monte Carlo (MCMC) approach in which we modify the assignment of each worker to a worker type and each job to a market in a random fashion and accept or reject each modification with a probability given as a function of the change in the likelihood. We repeat the procedure for multiple different starting values to reduce the chances of finding local maxima. We implement the procedure using a Python package called graph-tool. (\url{https://graph-tool.skewed.de/}. See \citet{Peixoto2014_efficient} for details.) Now that we have dealt with the issue of important worker and job characteristics being unobserved, we turn our attention to estimating counterfactuals for wage gap decompositions. 


\section{Wage gap decomposition}\label{sec:wage_gap_decomposition}

This section lays out the estimation strategies we use to decompose the Brazilian gender wage gap, while circumventing some of the issues associated with conventional decomposition methods. We decompose the gender wage gap into the quantities listed in equation (\ref{eq:overall_gap_decomp}): the composition component $E[Y_1(x_{ij}) | G_i = 1] - E[Y_1(x_{ij}) | G_i = 0]$ and the structural component  $E[Y_1(x_{ij}) - Y_0(x_{ij}) | G_i = 0]$. The quantity $E[Y_g(x_{ij})|G_i=g] = E[Y_{ij}|G_i=g]$, $g \in \{0,1\}$ can be consistently and straightforwardly estimated since it is directly observable. The challenge is estimating the counterfactual wage function $E[Y_1(x_{ij})|G_i=0]$, given that the potential outcome $Y_1(x_{ij})$ is not observed for female workers. Estimating $E[Y_1(x_{ij})|G_i=0]$ requires us to use data on male workers to estimate a relationship between observable characteristics $x_{ij}$ and male earnings $Y_1$ and then extrapolate this relationship to female workers. 
                
In this paper, we consider two approaches to estimating counterfactual wage functions. The first is the commonly-used Oaxaca-Blinder decomposition, which we henceforth refer to as OB \citep{oaxaca1973,blinder1973}. For the OB decomposition, we estimate two linear regressions --- one for the set of male workers and another for the set of female workers --- to estimate the functionals $Y_1(\cdot)$ and $Y_0(\cdot)$, respectively, as denoted in equation (\ref{eq:oaxaca_blinder_equations}). Values for $E[Y_g(x_{ij})|G_i=g]$ are obtained by averaging out the fitted values of the respective linear regressions. Estimates for the counterfactual $E[Y_1(x_{ij})|G_i=0]$ are obtained by using the coefficients from the linear regression fitted for males, $\hat\beta_{G=1}$, and multiplying them by the average female covariates, $\bar x_{G=0}$, as defined in equation (\ref{eq:oaxaca_blinder_equations}). This is equivalent to producing fitted values for the males' regression, while inputting females' covariates.

\begin{flalign}\label{eq:oaxaca_blinder_equations}
    &\text{OB regressions: } && Y_g(x_{ij}) = x_{ij} ^T \beta_{G=g} + \epsilon_{gij}, \qquad g \in \{0,1\} &&& \\				
    &\text{OB counterfactual estimate: }&& \widehat{ E[Y_1(x_{ij}) | G_i = 0]} := \bar x_{G=0} ^T \hat\beta_{G=1}, \quad \bar x_{G=0}:= \sum_{i | G_i=0} \frac{x_{ij}}{n} \nonumber &&&
\end{flalign}

As discussed in section \ref{sec:decomposition_motivation}, the OB decomposition has several important  limitations. Although highly tractable, OB imposes potentially restrictive assumptions on $Y_1(x_{ij})$. First, it assumes that its expectation is linear in $x_{ij}$. Although linear regressions allow for flexible transformations of its covariates, the functional form is still a somewhat arbitrary researcher choice. Second, by using a linear regression to estimate the potential outcome function, $Y_1(\cdot)$, as in equation (\ref{eq:oaxaca_blinder_equations}), it uses the same functional form to compute counterfactuals for all male workers. In other words, it imposes the same average returns to covariates for all workers, which would create biases in the counterfactual estimation if returns to worker characteristics are heterogeneous. The third limitation of the OB is related to the \textit{overlapping supports} assumption, also referred to as the \textit{common supports} assumption. This assumption imposes that the support of $x$ for one of the genders has to fully overlap with the support of $x$ for the other gender, and is imposed by almost all decomposition methods in economics (\citealp{fortinlemieuxfirpo2011}). The overlapping supports assumption is imposed to ensure that the counterfactual function $Y_1(x)$ estimated using male data, $x_{G_i=1}$, is only used to predict counterfactual earnings for females whose values of $x$ lie within the male support of $x$. When this condition is not satisfied in the data, observations that are outside of the common support are typically trimmed or given virtually zero weight in the estimation process, potentially eliminating significant numbers of workers from the analysis and making the analysis representative of only a subset of the population \citep{Modenesi2022}. This is particularly salient when $x$ lies in a high-dimensional space, as is the case in our application with high-dimensional worker types and markets. 

Our preferred decomposition strategy relies on matching male and female workers with similar observable characteristics and using matched workers of different genders as counterfactuals for each other. This approach was initially proposed by \citet{nopo2008} and was further extended by \citet{Modenesi2022}. Not only does this approach avoid the strong functional form assumptions made by OB, it includes a framework for handling a lack of common support. In this paper, we choose to use the original estimation strategy laid out by \citet{nopo2008}, given its tractability especially for a high-dimensional set of covariates like ours, and we refer to it as the matching decomposition henceforth.
            
The matching decomposition has two main components: (i) matching observations and (ii) relaxing the overlapping supports assumption. First, counterfactual female earnings $Y_1(x_{ij})|G_i=0$ --- what female workers would have earned if their gender were changed to male but nothing else about them changed --- are obtained by \textit{exact matching} each female to one or more male workers with similar observable characteristics and then taking a sample average of the matched males\footnote{In this paper we coarsened a few variables such as years of education and age, and we use the coarsened version of these variables instead to perform the exact matching. This serves the purpose of matching more individuals, giving more statistical power to the method, since workers with just e.g. 1 year difference in age, \textit{ceteris paribus}, are roughly the same in terms of productivity.}. This method for building counterfactuals is non-parametric, assuming no functional form for $Y_1(\cdot)$, it exerts no extrapolations out of the support of $x$ and it avoids using data from all workers to build counterfactuals for a specific worker. The matching decomposition handles the lack of common support issue by allowing unmatched workers, i.e. outside of the common support of $x$, to contribute to the overall observed gap. In the matching decomposition, we add two terms, $\Delta_M$ and $\Delta_F$, to the expression for the overall wage gap $\Delta$ in equation (\ref{eq:overall_gap}) which captures the contributions of unmatched male and female workers, respectively. The resulting expression is 

\begin{flalign}\label{eq:np_decomposition}
    \Delta =& E[Y_{ij} | G_i = 1] - E[Y_{ij} | G_i = 0] =: \Delta_X + \Delta_0 + \Delta_M + \Delta_F, 
\end{flalign}
where
\begin{flalign*}
    &\Delta_X := E \left[Y_{ij} | Matched, G_i= 1 \right] - E \left[Y_{1}(x_{ij}) | Matched, G_i= 0 \right]   \nonumber \\
    &\Delta_0 := E \left[Y_{ij} | Matched, G_i= 1 \right] - E \left[Y_{1}(x_{ij}) | Matched, G_i= 0 \right]   \nonumber \\
    &\Delta_M :=  \left\{ E \left[Y_{ij} | Unmatched, G_i= 1 \right] - E \left[Y_{ij} | Matched, G_i= 1 \right]  \right\} P \left( Unmatched | G_i= 1\right)  \nonumber \\
    &\Delta_F := \left\{ E \left[Y_{ij} | Matched, G_i= 0 \right] - E \left[Y_{ij} | Unmatched, G_i= 0 \right]  \right\} P \left( Unmatched | G_i= 0 \right)   \nonumber
\end{flalign*}

Notice that if all observations are matched the $\Delta_M$ and $\Delta_F$ terms vanish and this method collapses back to the original decomposition we have in equation (\ref{eq:overall_gap_decomp}). The terms $\Delta_X$ and $\Delta_0$ still have the same interpretation as discussed in Section \ref{sec:decomposition_motivation} --- composition and structural, respectively --- but now only similar workers of one gender are used to build counterfactuals for the other gender, using an agnostic functional form for the counterfactual function. The extra terms $\Delta_M$ and $\Delta_F$ measure the contribution of unmatched male and female workers to the overall observed gender gap. Each of them measures the difference between matched and unmatched workers of a given gender, weighted by the proportion of unmatched workers within that gender\footnote{Precise definitions of each of the terms in the NP decomposition can be found in the appendix section \ref{app:NP_decomp}}.   For example, if unmatched male workers have an average log wage that is 0.2 higher than the average log wage for matched male workers and 10\% of male workers are unmatched, then $\Delta_M = 0.2 \times 0.1 = 0.02$. 

To understand how the matching decomposition handles a lack of common support, consider male workers employed as professional football players. These workers will not be matched to female workers and therefore would be omitted from the analysis if we simply restrict it to the region of common support. However, the male workers do contribute meaningfully to the overall gender wage gap because they earn significantly more than the average female worker. The matching decomposition would handle this by including these workers in the $\Delta_M$ term. Intuitively, it would say that some of the gender wage gap can be decomposed within the region of common support, while some of it is explained by male workers outside the region of common support earning more than male workers within the region of common support, and similarly for female workers.

Our preferred specifications in this paper use the matching decomposition in conjunction with the latent skills and tasks clusters revealed by our network methodology developed in Section \ref{sec:model_and_sbm}. Since we define labor market gender discrimination as workers with similar skills performing similar tasks with similar productivity but being paid differently based on gender, our worker type--market clusters serve as natural cells within which workers are considered as equivalent in terms of productivity. With the matching decomposition we are able to ensure that only similar workers are used when estimating counterfactual earnings, mitigating counterfactual biases, and also avoid dropping unmatched workers from the estimation procedure as mentioned above. Although the original matching decomposition is not considered to be a ``detailed decomposition'' by the literature of decompositions in economics, in combination with our network clusters, it is possible to compute an economically principled distribution of the gender gap (and its components) for a vast amount of cells of workers in the labor market, mapping how discrimination is spread in different parts of the market. 


\section{Data}
\label{sec:data_ch2}

\subsection{Administrative Brazilian data}
We use the Brazilian linked employer-employee data set RAIS. The data contain detailed information on all employment contracts in the Brazilian formal sector, going back to the 1980s. The sample we work with includes all workers between the ages of 25 and 55 employed in the formal sector in the Rio de Janeiro metro area at least once between 2009 and 2018. These workers are defined as matching with the unemployment (or informal sector) in years we do not observe them. We also exclude the public sector because institutional barriers make flows between the Brazilian public and private sectors rare, as well as the military. Finally, we exclude the small number of jobs that do not pay workers on a monthly basis.

Our wage variable is the real hourly log wage in December, defined as total December earnings divided by hours worked. We deflate wages using the national inflation index. We exclude workers who were not employed for the entire month of December because we do not have accurate hours worked information for such workers. We define a job as an occupation-establishment pair. This implicitly assumes that all workers employed in the same occupation at the same establishment are performing approximately the same tasks. 

\begin{figure}
    \centering
    \caption{Distributions of Number of Matches Per Worker and Job}
    \begin{subfigure}{\textwidth}
        \centering
        \caption{Workers}
        \includegraphics[width=0.7\linewidth]{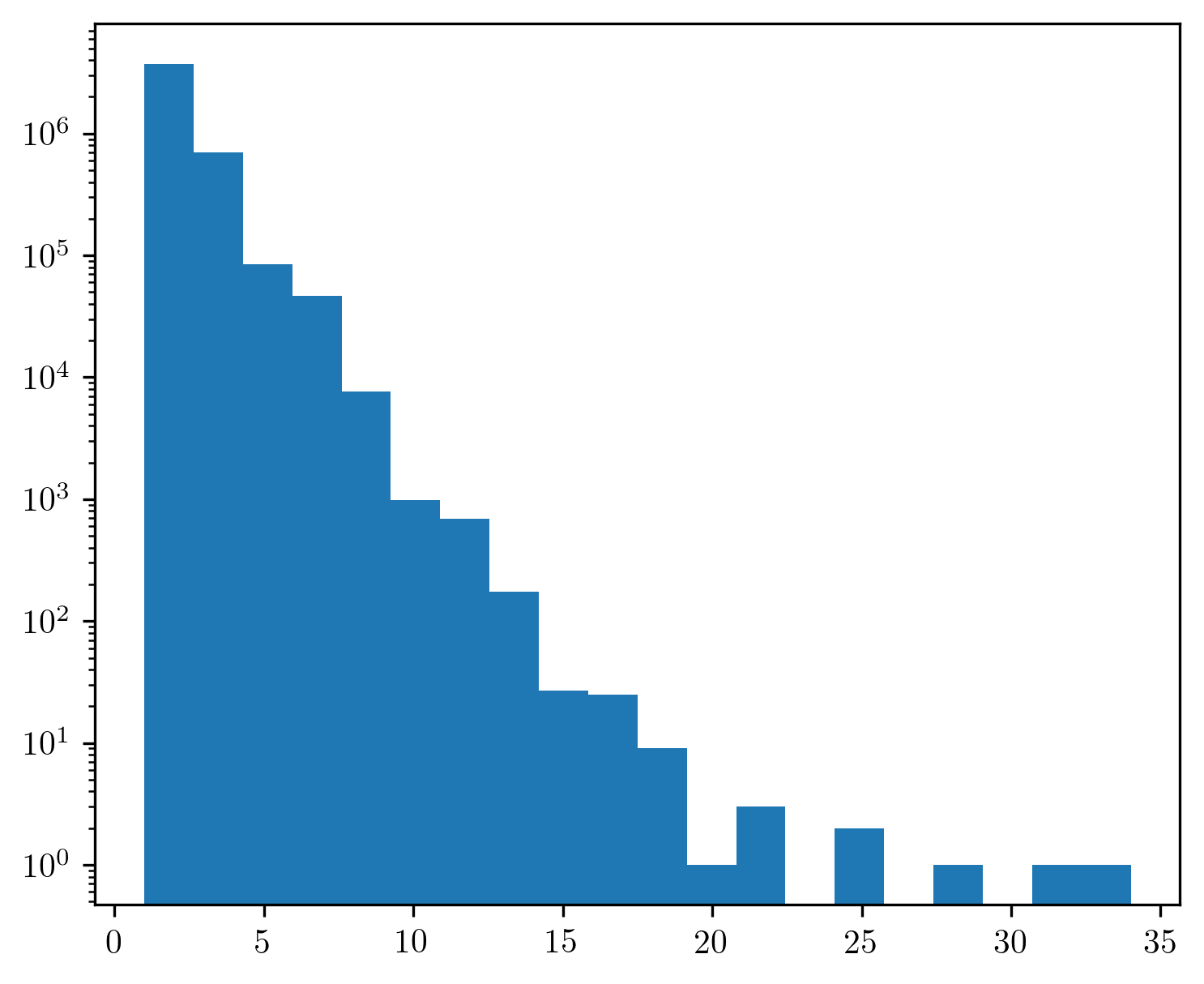}
        \label{fig:worker_degree_distribution_hist}
    \end{subfigure}
    \begin{subfigure}{\textwidth}
        \centering
        \caption{Jobs}
        \includegraphics[width=0.7\linewidth]{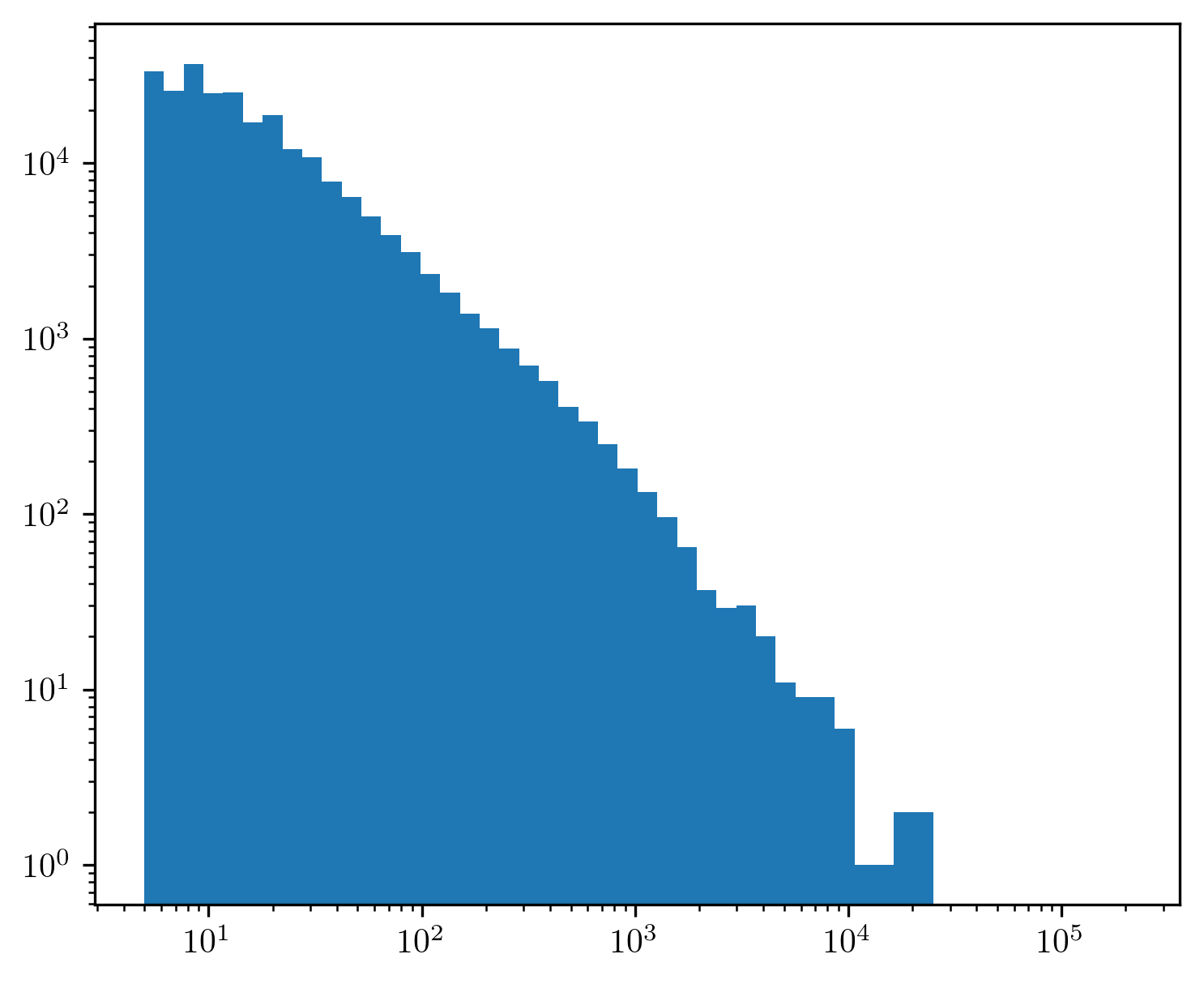}
        \label{fig:job_degree_distribution_hist}
    \end{subfigure}
    \label{fig:degree_distribution_hist}
    \footnotesize\flushleft \emph{Notes:} Figure presents histograms of the number of matches for workers and jobs, respectively. In network theory parlance, these are known as degree distributions. Vertical axes presented in log scale. Horizontal axis of bottom panel also presented in log scale. Number of matches per worker and job computed from the network of worker--job matches described in Section \ref{sec:data_ch2}.
\end{figure}

Our data contain 4,578,210 unique workers, 289,836 unique jobs, and 7,940,483 unique worker--job matches. The average worker matches with 1.73 jobs and the average job matches with 27.4 workers. 42\% of workers match with more than one job during our sample. Figure \ref{fig:degree_distribution_hist} presents histograms of the number of matches for workers and jobs, respectively. In network theory parlance, these are known as degree distributions. 

\begin{figure}
    \centering
    \caption{Worker Type ($\i$) and Market ($\g$) Size Distributions}
    \begin{subfigure}{\textwidth}
        \centering
        \caption{Number of Workers Per Worker Type ($\i$)}
        \includegraphics[width=0.7\linewidth]{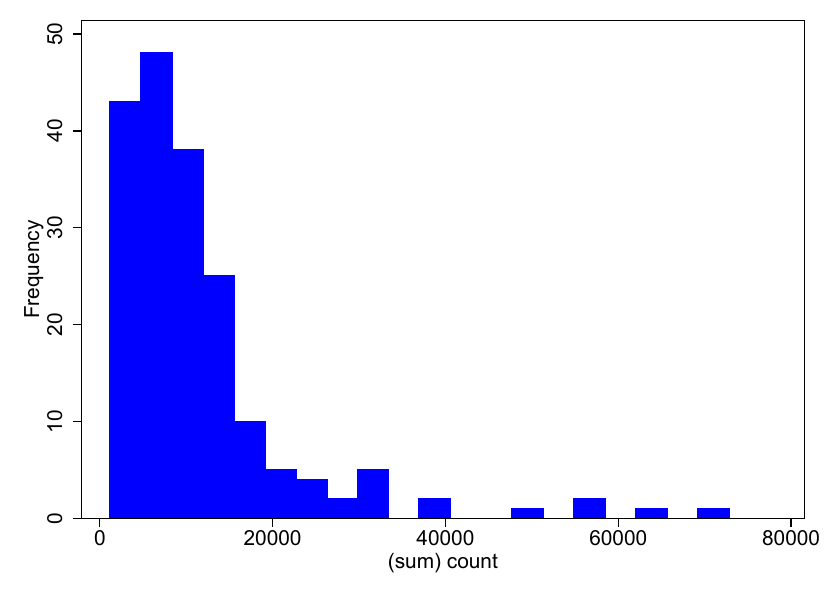}
        \label{fig:iota_size_distribution}
    \end{subfigure}
    \begin{subfigure}{\textwidth}
        \centering
        \caption{Number of Jobs Per Market ($\g$)}
        \includegraphics[width=0.7\linewidth]{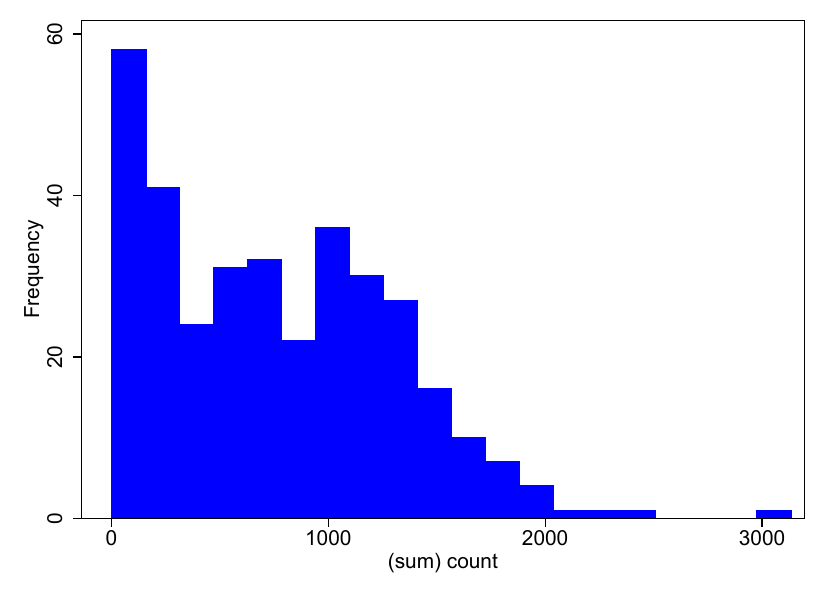}
        \label{fig:gamma_size_distribution}
    \end{subfigure}
    \label{fig:iota_gamma_size_distribution}
    \footnotesize\flushleft \emph{Notes:} Figure presents histograms of the number of workers per worker type $\i$ and jobs per market ($\g$). The units of analysis are worker types in the upper panel and markets in the lower panel. Computed using assignments of workers to worker types and jobs to markets as described in Section \ref{sec:bisbm}.
\end{figure}

Our network-based classification algorithm identifies 187 worker types ($\i$) and 341 markets ($\g$). Figure \ref{fig:iota_gamma_size_distribution} presents histograms of the number of workers per worker type and jobs per market. The average worker belongs to a worker type with 20,896 workers and the median worker belongs to a worker type with 14,211 workers. The average job belongs to a market with 1,156 jobs and the median job belongs to a market with 1,127 jobs.

\section{Results}
\label{sec:results}

\subsection{Aggregate wage gap decomposition}

\label{sec:aggregate_decomposition}

Table \ref{table:aggregated_decompostions} presents the results of performing gender wage decompositions using each of our two methods: OB and matching. For each method, we have three specifications. The first, presented in columns (1) and (4), estimates counterfactual earnings distributions using a standard set of observable characteristics: experience, education, race, industry and union status. The second, presented in columns (2) and (5), estimates counterfactual earnings distributions using the worker types and markets identified by the SBM. The third specification, presented in columns (3) and (6) uses both standard observable characteristics and worker types and markets. The first row of each column presents the overall wage gap: the average male worker earns 16.7 percent more than the average female worker in our sample. The second row presents the wage gap that would exist if male and female workers with the same productivity were paid equivalently but the observed differences between the distributions of male and female productivity --- as proxied by observable characteristics and/or worker types and markets --- remained, the \textit{composition} component. The third row presents the wage gap that would exist if male and female workers had identical productivity distributions, but the observed earnings differences conditional on productivity remained, the \textit{structural} component. The fourth and fifth rows present the wage gap explained by male and female workers outside the region of common support, respectively.  For the OB method the composition and structural components add up to the overall wage gap; for the matching method the overall wage gap equals the sum of the composition and structural components and the components due to a lack of common support.

The qualitative stories told by both the OB method and the matching method are similar. When we define counterfactual earnings using observable characteristics (columns 1 and 4), we find that if male and female workers with the same productivity were paid similarly, then female workers would significantly outearn male workers (structural effect): by 12.7\% using the OB method and 8.8\% using the matching method. By contrast, female workers would be paid significantly less if they possessed the male's productivity distribution (composition effect):  29.4\% less using the OB method and 25.6\% less using the matching method. When we define counterfactuals using worker types and markets instead of observable characteristics (columns 2 and 5) we find that the wage gap would nearly disappear if male and female workers with the same productivity were paid similarly. By contrast, the wage gap that would exist if male and female workers had the same productivity distribution --- 17.9\% according to OB and 17.8\% according to matching --- is almost equal to the overall wage gap of 16.7\%. In other words, when we compute counterfactuals using worker types and markets we find that differential pay for similar productivity explains roughly the entire gender wage gap. This tells us that the results of gender wage gap decompositions are highly sensitive to the way in which we define counterfactuals. If, as we argue, worker types and markets do a better job of capturing the latent productivity of worker--job matches than do standard observable characteristics, then these results imply that gender wage gaps are almost entirely due to similarly productive male and female workers being paid differently, not male and female workers having different productivity distributions. 

Columns (3) and (6) of Table \ref{table:aggregated_decompostions} use both observable characteristics and worker types and markets to form counterfactuals for the gender wage gap decompositions. The OB method finds that female workers have covariates that would imply that they would outearn male workers if equally productive workers were paid equivalently, similar to the findings when we included only observable characteristics, not worker and job types, in column (1). By contrast, the matching method finds that male workers' covariates imply 3.4\% higher earnings than female workers' covariates and that male workers are paid 18.5\% more than similarly productive female workers. Why do we observe a discrepancy between the OB and matching methods once we include observable characteristics and worker types and markets? The answer lies in the final two rows of Table \ref{table:aggregated_decompostions}, which present the fraction of male and female workers, respectively, for whom we are unable to find a counterfactual. Once we try to match workers on such a large set of variables, many workers are unable to be matched, and a significant part of the gender wage gap occurs among such workers. The matching method allows us to take this into account, while the OB method simply makes a linear extrapolation. However, a linear extrapolation outside the region of common support is likely to lead to incorrect inferences. Furthermore, the fact that the matching estimator yields similar results when we use worker types and markets as it does when we use worker types, markets, and other observable characteristics, but not when we use other observables alone, implies that worker types and markets capture significant determinants of productivity, and omitting them leads to incorrect inferences. This highlights the importance of using a sufficiently set of worker characteristics when estimating counterfactuals, and our method for identifying previously unobserved heterogeneity enhances our ability to do so. All of the results presented in this section correspond to the aggregate gender wage gap. In the next section, we consider heterogeneity in wage gaps within different subsets of the labor market.

\begin{table}[htbp!]
    \centering
    \caption{Gap decomposition using Oaxaca-Blinder vs Matching}
    \begin{tabular}{lcccccc}
        \toprule
        & \multicolumn{3}{c}{Oaxaca-Blinder}   & \multicolumn{3}{c}{Matching} \\ \cline{2-7} 
        & Observables & $\iota\times\gamma$ & Full model & Observables      & $\iota\times\gamma$      & Full model     \\
        & (1) & (2)   & (3)   & (4) &     (5)     & (6) \\
        \midrule
        Gap                   & 0.167      & 0.167      & 0.167      & 0.167           & 0.167           & 0.167          \\
        \hspace{.3cm} Composition           & -0.127     & -0.011     & -0.084     & -0.088          & -0.006          & 0.034          \\
        \hspace{.3cm} structural             & 0.294      & 0.179      & 0.250      & 0.256           & 0.178           & 0.185          \\
        \hspace{.3cm} Males unmatched       & -          & -          & -          & 0.000           & -0.005          & -0.076         \\
        \hspace{.3cm} Females unmatched     & -          & -          & -          & 0.000           & 0.000           & 0.024          \\
        \midrule
        \% of males matched   & -          & -          & -          & 1.00            & 0.98            & 0.57           \\
        \% of females matched & -          & -          & -          & 1.00            & 0.99            & 0.74
        \\
        \bottomrule
    \end{tabular} \label{table:aggregated_decompostions}
    \flushleft \footnotesize \emph{Notes:} All coefficients significant at at least the 1\% level.
\end{table}


\subsection{Wage gaps within worker type--market cells}

An appealing feature of our worker types and markets is that they allow us to further decompose gender wage gaps and identify heterogeneity in gender wage gaps across the labor market. We do so by computing overall wage gaps, $\Delta$, and then decomposing them following the matching decomposition, \textit{within} each worker type--market cell. 

\begin{figure}[htbp!]
    \centering
    \caption{Distribution of Components of Overall Wage Gap, Disaggregated}
    \includegraphics[width=1.0\linewidth]{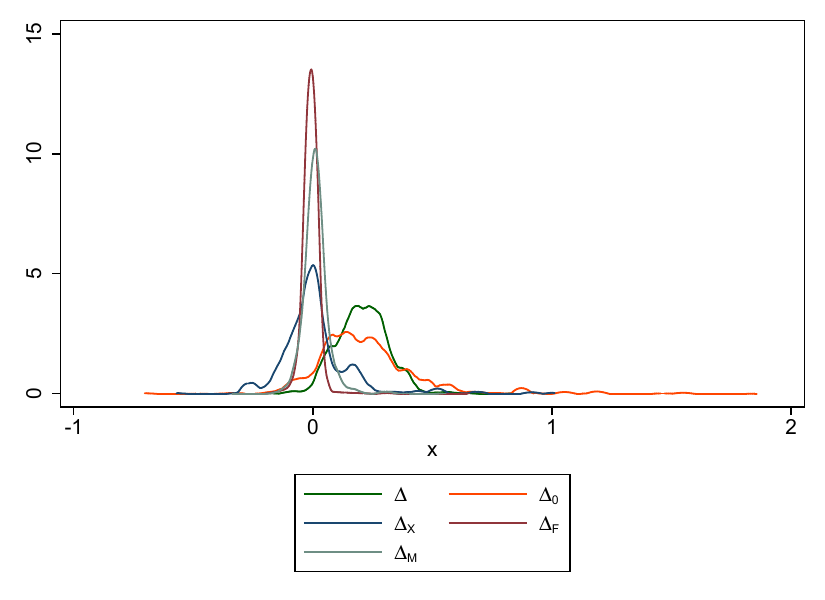}
    \label{fig:kdensity_wDs_g1000}
    \footnotesize\flushleft \emph{Notes:} All values measured as differences between male and female log wages. Weighted by number of workers per worker type--market cell. Worker type--market cells with fewer than 1000 workers dropped to remove outliers and improve visual clarity.
\end{figure}

For each worker type--market cell we decompose the overall wage gap (Row 1 of Table \ref{table:aggregated_decompostions}) into its four components: composition, structural, males unmatched, and females unmatched (Rows 2--5 of Table \ref{table:aggregated_decompostions}). Figure \ref{fig:kdensity_wDs_g1000} presents kernel density plots of the resulting distributions of overall wage gaps and their four components. Several clear patterns emerge. First, the overall wage gaps $\Delta$ are almost universally positive, meaning that male workers outearning their female counterparts is a widespread phenomenon. Specifically, 91\% of workers are in clusters where males outearn females. Second, the distribution of the structural component, $\Delta_0$, is similar to the distribution of the overall wage gap. This suggests that the result from the aggregate decomposition in Section \ref{sec:aggregate_decomposition} that almost the entire overall gender wage gap is explained by the structural component holds within worker type--market cells as well. The fact that the structural component roughly coincides with the overall wage gap implies that the other three components --- composition, males outside the common support, and females outside the common support --- must contribute relatively little to the overall gender wage gap, which is confirmed by the fact that the distributions for these three components are centered close to zero and have low variances. We present the same results quantitatively in Table \ref{table:sumstats}. Together, these results tell us that while there is significant variability in gender wage gaps across different worker type--market pairs, the overall qualitative pattern of male workers outearning their female counterparts, and almost all of this gap being explained by differential returns to the same skills rather than different skills, is true in the disaggregated results as well as the aggregated results. 

\begin{table}[htbp!]
    \centering
    \caption{Summary Statistics of Components of Overall Wage Gap, Disaggregated}
    {
\def\sym#1{\ifmmode^{#1}\else\(^{#1}\)\fi}
\begin{tabular}{l*{1}{ccccc}}
\hline\hline
                    &\multicolumn{5}{c}{}                                            \\
                    &        mean&          sd&         min&         max&       count\\
\hline
$\Delta$            &       0.215&       0.240&      -1.183&       6.228&     4791014\\
\hline
$\Delta_0$          &       0.196&       0.172&      -2.506&       9.384&     4791014\\
$\Delta_M$          &       0.016&       0.134&      -3.577&       3.448&     4783255\\
$\Delta_F$          &      -0.011&       0.116&      -2.632&       4.684&     4724863\\
$\Delta_X$          &       0.013&       0.153&      -1.150&       2.418&     4791014\\
\hline
Frac. Male Workers Matched&       0.766&       0.238&       0.004&       1.000&     4791014\\
Frac. Female Workers Matched&       0.875&       0.199&       0.008&       1.000&     4791014\\
Frac. Workers that Are Male&       0.617&       0.162&       0.037&       0.999&     4791014\\
\hline\hline
\end{tabular}
}
    \label{table:sumstats}
\end{table}

\clearpage


\section{Conclusion}
\label{sec:conclusion_ch2}

In this paper we reconsider the wage gap decomposition literature and make three key contributions. First, we propose a new method for identifying unobserved determinants of workers’ earnings from the information revealed by detailed data on worker–job matching patterns. The method builds on \citet{FogelModenesi2021} and provides a blueprint for incorporating observable variables into the clustering algorithm, while also relaxing the assumption of perfect competition in labor markets. Second, we non-parametrically estimate counterfactual wage functions for male and female workers and use them to decompose gender wage gaps into a \textit{composition} component in which male and female workers earn different wages because they possess different skills and perform different tasks, and a \textit{structural} component in which male and female workers who possess similar skills and perform similar tasks nonetheless earn different wages. Third, we address the issue of male workers' observables characteristics falling outside the support of female workers' observable characteristics, and vice versa, by augmenting the wage decomposition with components attributable to male and female workers, respectively, outside the region of common support.

We apply these methods to Brazilian administrative data and find that almost the entire gender wage gap is attributable to male and female workers who possess similar skills and perform similar tasks being paid differently. This is true at the aggregate level, and remains true when we perform wage decompositions within each worker type--market cell, indicating that this is a widespread phenomenon, not one driven by large wage differentials in small subsets of the labor market. We find that wage decompositions based on standard observable variables suffer from omitted variable bias, emphasizing the need for detailed worker and job characteristics in the form of worker types and markets. We find that wage decompositions based on linear regressions yield similar findings to those based on matching when a lack of common support is not an issue, however when male and female workers' characteristics do not share a common support the matching estimator with corrections for a lack of common support outperforms alternatives. 

While this paper focuses on gender wage gaps, the methods are applicable to other wage gaps, for instance race. Moreover, our strategy for using worker--job matching patterns to control for previously-unobserved, but potentially confounding, covariates may be applied in a wide variety of contexts.

	
\clearpage
\bibliographystyle{aer}	
\bibliography{References.bib}

@article{AutorLevyMurnane2003,
	title        = {{The Skill Content of Recent Technological Change: An Empirical Exploration}},
	author       = {David H. Autor and Frank Levy and Richard J. Murnane},
	year         = 2003,
	month        = {},
	journal      = {The Quarterly Journal of Economics},
	volume       = 118,
	number       = 4,
	pages        = {1279--1333},
	doi          = {},
	url          = {https://ideas.repec.org/a/oup/qjecon/v118y2003i4p1279-1333..html}
}

@book{Train2009,
	title        = {Discrete choice methods with simulation},
	author       = {Train, Kenneth E},
	year         = 2010,
	publisher    = {Cambridge university press}
}

@article{Mcfadden1978,
	title        = {Modeling the choice of residential location},
	author       = {McFadden, Daniel},
	year         = 1978,
	journal      = {Transportation Research Record},
	number       = 673
}

@article{Nimczik2018,
	title        = {Job Mobility Networks and Endogenous Labor Markets},
	author       = {Nimczik, Jan Sebastian},
	year         = 2018,
	url          = {https://drive.google.com/file/d/0B0-Gl5fHwGOBOXlOVHNWTktCODQ/view}
}

@article{CardCardosoHeiningKline2018,
	title        = {Firms and Labor Market Inequality: Evidence and Some Theory},
	author       = {Card, David and Cardoso, Ana Rute and Heining, Joerg and Kline, Patrick},
	year         = 2018,
	journal      = {Journal of Labor Economics},
	volume       = 36,
	number       = {S1},
	pages        = {S13-S70},
	doi          = {10.1086/694153},
	url          = {https://doi.org/10.1086/694153},
	eprint       = {https://doi.org/10.1086/694153}
}

@article{Tan2018,
	title        = {Multidimensional heterogeneity and matching in a frictional labor market - An application to polarization},
	author       = {Tan, Joanne},
	year         = 2018
}

@article{Lindenlaub2017,
	title        = {Sorting multidimensional types: Theory and application},
	author       = {Lindenlaub, Ilse},
	year         = 2017,
	journal      = {The Review of Economic Studies},
	publisher    = {Oxford University Press},
	volume       = 84,
	number       = 2,
	pages        = {718--789}
}

@article{Peixoto2014_efficient,
	title        = {Efficient Monte Carlo and greedy heuristic for the inference of stochastic block models},
	author       = {Peixoto, Tiago P},
	year         = 2014,
	journal      = {Physical Review E},
	publisher    = {APS},
	volume       = 89,
	number       = 1,
	pages        = {012804}
}

@article{LarremoreClausetJacobs2014,
	title        = {Efficiently inferring community structure in bipartite networks},
	author       = {Larremore, Daniel B and Clauset, Aaron and Jacobs, Abigail Z},
	year         = 2014,
	journal      = {Physical Review E},
	publisher    = {APS},
	volume       = 90,
	number       = 1,
	pages        = {012805}
}

@article{Sorkin2018,
	title        = {Ranking firms using revealed preference},
	author       = {Sorkin, Isaac},
	year         = 2018,
	journal      = {The quarterly journal of economics},
	publisher    = {Oxford University Press},
	volume       = 133,
	number       = 3,
	pages        = {1331--1393}
}

@techreport{JaroschNimczikSorkin2019,
	title        = {Granular search, market structure, and wages},
	author       = {Jarosch, Gregor and Nimczik, Jan Sebastian and Sorkin, Isaac},
	year         = 2019,
	institution  = {National Bureau of Economic Research}
}

@article{Peixoto2019,
	title        = {Bayesian stochastic blockmodeling},
	author       = {Peixoto, Tiago P},
	year         = 2019,
	journal      = {Advances in network clustering and blockmodeling},
	publisher    = {Wiley Online Library},
	pages        = {289--332}
}

@article{Autor2013,
	title        = {The `task approach' to labor markets: an overview},
	author       = {Autor, David H},
	year         = 2013,
	publisher    = {MIT Department of Economics Working Paper}
}

@article{AcemogluAutor2011,
	title        = {Skills, tasks and technologies: Implications for employment and earnings},
	author       = {Acemoglu, Daron and Autor, David},
	year         = 2011,
	booktitle    = {Handbook of labor economics},
	publisher    = {Elsevier},
	volume       = 4,
	pages        = {1043--1171}
}

@article{Peixoto2017,
	title        = {Nonparametric Bayesian inference of the microcanonical stochastic block model},
	author       = {Peixoto, Tiago P},
	year         = 2017,
	journal      = {Physical Review E},
	publisher    = {APS},
	volume       = 95,
	number       = 1,
	pages        = {012317}
}

@article{Kantenga2018,
	title        = {The effect of job-polarizing skill demands on the US wage structure},
	author       = {Kantenga, Kory},
	year         = 2018,
	publisher    = {European University Institute}
}

@article{Roy1951,
	title        = {Some thoughts on the distribution of earnings},
	author       = {Roy, Andrew Donald},
	year         = 1951,
	journal      = {Oxford economic papers},
	publisher    = {JSTOR},
	volume       = 3,
	number       = 2,
	pages        = {135--146}
}

@techreport{garcianoposalardi2009,
	title        = {{Gender and Racial Wage Gaps in Brazil 1996-2006: Evidence Using a Matching Comparisons Approach}},
	author       = {Luana Marquez Garcia and Hugo \~{N}opo and Paola Salardi},
	year         = 2009,
	month        = May,
	number       = 4626,
	doi          = {},
	url          = {https://ideas.repec.org/p/idb/wpaper/4626.html},
	institution  = {Inter-American Development Bank, Research Department},
	type         = {Research Department Publications}
}

@article{morelloanjolim2021,
	title        = {Gender wage discrimination in Brazil from 1996 to 2015: A matching analysis},
	author       = {Thiago Morello and Jacqueline Anjolim},
	year         = 2021,
	journal      = {EconomiA},
	doi          = {https://doi.org/10.1016/j.econ.2021.03.002},
	issn         = {1517-7580},
	url          = {https://www.sciencedirect.com/science/article/pii/S1517758021000138}
}

@techreport{hurstrubinsteinshimizu2021,
	title        = {Task-Based Discrimination},
	author       = {Hurst, Erik and Rubinstein, Yona and Shimizu, Kazuatsu},
	year         = 2021,
	month        = {July},
	series       = {Working Paper Series},
	number       = 29022,
	doi          = {10.3386/w29022},
	url          = {http://www.nber.org/papers/w29022},
	institution  = {National Bureau of Economic Research},
	type         = {Working Paper}
}

@article{cardcardosokline2015,
	title        = {{ Bargaining, Sorting, and the Gender Wage Gap: Quantifying the Impact of Firms on the Relative Pay of Women *}},
	author       = {Card, David and Cardoso, Ana Rute and Kline, Patrick},
	year         = 2015,
	month        = 10,
	journal      = {The Quarterly Journal of Economics},
	volume       = 131,
	number       = 2,
	pages        = {633--686},
	doi          = {10.1093/qje/qjv038},
	issn         = {0033-5533},
	url          = {https://doi.org/10.1093/qje/qjv038},
	eprint       = {https://academic.oup.com/qje/article-pdf/131/2/633/30636241/qjv038.pdf}
}

@article{goldin2014,
	title        = {A Grand Gender Convergence: Its Last Chapter},
	author       = {Goldin, Claudia},
	year         = 2014,
	month        = {April},
	journal      = {American Economic Review},
	volume       = 104,
	number       = 4,
	pages        = {1091--1119},
	doi          = {10.1257/aer.104.4.1091},
	url          = {https://www.aeaweb.org/articles?id=10.1257/aer.104.4.1091}
}

@article{barskyboundcharleslupton2002,
	title        = {Accounting for the Black-White Wealth Gap: A Nonparametric Approach},
	author       = {Robert Barsky and John Bound and Kerwin Kofi Charles and Joseph P. Lupton},
	year         = 2002,
	journal      = {Journal of the American Statistical Association},
	publisher    = {[American Statistical Association, Taylor & Francis, Ltd.]},
	volume       = 97,
	number       = 459,
	pages        = {663--673},
	issn         = {01621459},
	url          = {http://www.jstor.org/stable/3085702}
}

@article{firpofortinlemieux2018,
	title        = {{Decomposing Wage Distributions Using Recentered Influence Function Regressions}},
	author       = {Sergio P. Firpo and Nicole M. Fortin and Thomas Lemieux},
	year         = 2018,
	month        = {May},
	journal      = {Econometrics},
	volume       = 6,
	number       = 2,
	pages        = {1--40},
	doi          = {},
	url          = {https://ideas.repec.org/a/gam/jecnmx/v6y2018i2p28-d149033.html}
}

@techreport{gerardlagosseverninicard2018,
	title        = {Assortative Matching or Exclusionary Hiring? The Impact of Firm Policies on Racial Wage Differences in Brazil},
	author       = {Gerard, François and Lagos, Lorenzo and Severnini, Edson and Card, David},
	year         = 2018,
	month        = {October},
	series       = {Working Paper Series},
	number       = 25176,
	doi          = {10.3386/w25176},
	url          = {http://www.nber.org/papers/w25176},
	institution  = {National Bureau of Economic Research},
	type         = {Working Paper}
}

@article{nopo2008,
	title        = {Matching as a Tool to Decompose Wage Gaps},
	author       = {Hugo {\~N}opo},
	year         = 2008,
	journal      = {The Review of Economics and Statistics},
	publisher    = {The MIT Press},
	volume       = 90,
	number       = 2,
	pages        = {290--299},
	issn         = {00346535, 15309142},
	url          = {http://www.jstor.org/stable/40043147}
}

@article{blinder1973,
	title        = {Wage Discrimination: Reduced Form and Structural Estimates},
	author       = {Alan S. Blinder},
	year         = 1973,
	journal      = {The Journal of Human Resources},
	publisher    = {[University of Wisconsin Press, Board of Regents of the University of Wisconsin System]},
	volume       = 8,
	number       = 4,
	pages        = {436--455},
	issn         = {0022166X},
	url          = {http://www.jstor.org/stable/144855}
}

@article{oaxaca1973,
	title        = {Male-Female Wage Differentials in Urban Labor Markets},
	author       = {Ronald Oaxaca},
	year         = 1973,
	journal      = {International Economic Review},
	publisher    = {[Economics Department of the University of Pennsylvania, Wiley, Institute of Social and Economic Research, Osaka University]},
	volume       = 14,
	number       = 3,
	pages        = {693--709},
	issn         = {00206598, 14682354},
	url          = {http://www.jstor.org/stable/2525981}
}

@incollection{fortinlemieuxfirpo2011,
	title        = {Chapter 1 - Decomposition Methods in Economics},
	author       = {Nicole Fortin and Thomas Lemieux and Sergio Firpo},
	year         = 2011,
	publisher    = {Elsevier},
	series       = {Handbook of Labor Economics},
	volume       = 4,
	pages        = {1--102},
	doi          = {https://doi.org/10.1016/S0169-7218(11)00407-2},
	issn         = {1573-4463},
	url          = {https://www.sciencedirect.com/science/article/pii/S0169721811004072},
	editor       = {Orley Ashenfelter and David Card}
}

@article{dinardofortinlemieux1996,
	title        = {Labor Market Institutions and the Distribution of Wages, 1973-1992: A Semiparametric Approach},
	author       = {John DiNardo and Nicole M. Fortin and Thomas Lemieux},
	year         = 1996,
	journal      = {Econometrica},
	publisher    = {[Wiley, Econometric Society]},
	volume       = 64,
	number       = 5,
	pages        = {1001--1044},
	issn         = {00129682, 14680262},
	url          = {http://www.jstor.org/stable/2171954}
}

@article{chernozhukovfernandezvalmelly2013,
	title        = {Inference on Counterfactual Distributions},
	author       = {Chernozhukov, Victor and Fernández-Val, Iván and Melly, Blaise},
	year         = 2013,
	journal      = {Econometrica},
	volume       = 81,
	number       = 6,
	pages        = {2205--2268},
	doi          = {https://doi.org/10.3982/ECTA10582},
	url          = {https://onlinelibrary.wiley.com/doi/abs/10.3982/ECTA10582},
	eprint       = {https://onlinelibrary.wiley.com/doi/pdf/10.3982/ECTA10582}
}

@article{peixoto2018,
	title        = {Nonparametric weighted stochastic block models},
	author       = {Peixoto, Tiago P.},
	year         = 2018,
	month        = {Jan},
	journal      = {Phys. Rev. E},
	publisher    = {American Physical Society},
	volume       = 97,
	pages        = {012306},
	doi          = {10.1103/PhysRevE.97.012306},
	url          = {https://link.aps.org/doi/10.1103/PhysRevE.97.012306},
	issue        = 1,
	numpages     = 17
}

@article{KarrerNewman2011,
	title        = {Stochastic blockmodels and community structure in networks},
	author       = {Karrer, Brian and Newman, Mark EJ},
	year         = 2011,
	journal      = {Physical review E},
	publisher    = {APS},
	volume       = 83,
	number       = 1,
	pages        = {016107}
}

@misc{FogelModenesi2021,
	title        = {What is a Labor Market? Classifying Workers and Jobs Using Network Theory},
	author       = {Jamie Fogel and Bernardo Modenesi},
	year         = 2023,
	eprint       = {2311.00777},
	archiveprefix = {arXiv},
	primaryclass = {econ.GN}
}

@article{Modenesi2022,
	title        = {Advancing Distribution Decomposition Methods Beyond Common Supports: Applications to Racial Wealth Disparities},
	author       = {Bernardo Modenesi},
	year         = 2022
}


\clearpage
\appendix
\appendixpage

\section{Nested Logit Choice Probability}\label{app:ch2:choice-prob}

According to \citet{Train2009}, and originally developed by \citet{Mcfadden1978}, maximizing the utility choosing $j$, which is nested within a group $\g$ 
\begin{equation}\label{app:ch2:eq:trainmaximization}
    j^* := \arg\max_{j} \quad  W_\g + Y_j + \varepsilon_j
\end{equation}
with $\ve_j \sim NestedLogit(\nu_{\g})$ results in the following choice probability:
\begin{flalign*}
    P(j = j^*) 			&= P(\text{Choose } \g) P(j = j^* | \g)  \\
    &= \frac{\exp(W_\g + \nu_\g I_{\g})}{\sum_{\g} \exp(W_\g + \nu_\g I_{\g})}  \frac{\exp(Y_j)^\frac{1}{\nu_\g}}{\sum_{j \in \g} \exp(Y_j)^\frac{1}{\nu_\g}} 
    \end{flalign*}

where $I_\g = \log \left( \sum_{j \in \g} \exp(Y_j)^\frac{1}{\nu_\g} \right)$. \\

Our problem is similar, with workers choosing job $j$ within a market $\g$ in order to maximize the sum of log earnings $\log (\psi_{\ig} w_j^g)$  and an idiosyncratic preference for job $j$, $\ve_{i j}^g$:
\begin{flalign}\label{app:ch2:eq:ourmaximization}
    j^{*} 		=& \arg\max_{j} \quad \log (\psi_{\ig} w_j^g) + \ve_{i j}^g.  
\end{flalign}
We also assume that $\ve_{i j}^g \sim NestedLogit(\nu_{\g}^g)$. One of the differences from our setup to what is covered by \citet{Train2009} is that we add extra worker indexes $\i$ for her/his skills and $g$ for her gender and we condition our probabilities on knowing $\i$ and $g$. Notice that when comparing equations \ref{app:ch2:eq:trainmaximization} and \ref{app:ch2:eq:ourmaximization}, $W_\g = 0$ and $Y_j = \log (\psi_{\ig} w_j^g)$, which results in the following choice probabilities:
\begin{flalign*}
    P(j = j^* | j \in \g, i \in \i, g) 	&= P(\g = \g^* | i \in \i, j\in\g,g) P(j = j^* | i \in \i, j \in \g,\g = \g^*, g)  \\
    &= \frac{\exp(\nu_{\g}^g I_{\ig}^g)}{\sum_{\g} \exp(\nu_{\g}^g I_{\ig}^g)}  \frac{\exp(\log(\psi_{\ig} w_j^g))^\frac{1}{\nu_{\g}^g}}{\sum_{j \in \g} \exp(\log(\psi_{\ig} w_j^g))^\frac{1}{\nu_{\g}^g}} \quad \text{\tiny(plugging objects in)}\\
    &= \frac{\exp(I_{\ig}^g)^{\nu_{\g}^g}}{\sum_{\g} \exp(I_{\ig}^g)^{\nu_{\g}^g}}  \frac{(\psi_{\ig} w_j^g)^\frac{1}{\nu_{\g}^g}}{\sum_{j \in \g} (\psi_{\ig} w_j^g)^\frac{1}{\nu_{\g}^g}} \quad \text{\tiny(similar to equation \ref{eq_worker_choice})} \\
    &= \frac{\exp(I_{\ig}^g)^{\nu_{\g}^g}}{\sum_{\g} \exp(I_{\ig}^g)^{\nu_{\g}^g}}  \frac{(\psi_{\ig} w_j^g)^\frac{1}{\nu_{\g}^g}}{\exp(I_{\ig}^g)} \quad \text{\tiny(by definition of $I_{\ig}^g$)} \\
    &= \underset{\underset{\i-\g-g \text{ component}\quad}{\underbrace{=: \Omega_{\ig}^g}}}{\underbrace{\frac{\exp(I_{\ig}^g)^{\nu_{\g}^g-1}}{\sum_{\g} \exp(I_{\ig}^g)^{\nu_{\g}^g}}  \psi_{\ig}^\frac{1}{\nu_{\g}^g} }}   \underset{\underset{\quad j-g \text{ component}}{\underbrace{=: d_j^g  }}}{\underbrace{\vphantom{\frac{\exp(I_{\ig}^g)^{\nu_{\g}^g-1}}{\sum_{\g} \exp(I_{\ig}^g)^{\nu_{\g}^g}}  \psi_{\ig}^\frac{1}{\nu_{\g}^g}} (w_j^g)^\frac{1}{\nu_{\g}^g} }} \quad \text{\tiny(similar to equation \ref{eq_worker_choice_separation})} \\
    \end{flalign*}

where $I_{\ig}^g = \log \left[ \sum_{j \in \g} \exp(\log(\psi_{\ig} w_j^g))^\frac{1}{\nu_{\g}^g} \right] = \log \left[ \sum_{j \in \g} (\psi_{\ig} w_j^g)^\frac{1}{\nu_{\g}^g} \right]$.

\clearpage

\section{Terms in the NP decomposition}\label{app:NP_decomp}
The terms in the NP decomposition from equation \ref{eq:np_decomposition} can be more formally defined as follows:

\begin{flalign}
    &\Delta_M 	:=  \left[ \int_{\bar S_F} Y_{1}(x) \frac{dF_M(x)}{\mu_M(\bar S_F)} - \int_{S_F} Y_{1}(x) \frac{dF_M(x)}{\mu_M(S_F)}  \right] \mu_M(\bar S_F) && \\
    &\Delta_X 	:= \int_{S_M \cap S_F} Y_{1}(x) \left[\frac{dF_M(x)}{\mu_M(S_F)} - \frac{dF_F(x)}{\mu_F(S_M)}  \right] && \nonumber \\
    &\Delta_0 	:= \int_{S_M \cap S_F} \left[Y_{1}(x) - Y_{0}(x) \right] \frac{dF_F(x)}{\mu_F(S_M)} &&  \nonumber  \\
    &\Delta_F 	:= \left[ \int_{S_M} Y_{0}(x) \frac{dF_F(x)}{\mu_F(S_M)} - \int_{\bar S_M} Y_{0}(x) \frac{dF_F(x)}{\mu_F(\bar S_M)}\right] \mu_F(\bar S_M) 	&&   \nonumber			
\end{flalign}

where: $F_M(x)$ and $F_F(x)$ denote the distributions of $x$ for both males and females, respectively; $\mu_M$ and $\mu_F$ measure the proportions of males and females over regions of the supports of $x$; and the support of $x$ for a gender $g$, $supp_(X_g)$, is partitioned as $supp_(X_g) := S_g \cup \bar S_g$, with $S_g \cap \bar S_g = \emptyset $, for $g \in \{ F,M \}$.


\clearpage

\section{Proof that $A_{ij}$ follows a Poisson distribution}

\label{app:poisson_proof}

If an individual worker $i$ only searched for a job once, then the probability of worker $i$ matching with job $j$ would be equal to $\P_{ij} = \mathcal{P}_{\ig} d_j $ and $A_{ij}$ would follow a Bernoulli distribution: 
\[ A_{ij} \sim Bernoulli(\mathcal{P}_{\ig} d_j ). \]
However, since worker $i$ searches for jobs $c_i\equiv \sum_{t=1}^T c_{it}$ times, $A_{ij}$ is actually the sum of $c_i$ Bernoulli random variables, and is therefore a Binomial random variable. Conditional on knowing $c_i$, 
\[ A_{ij}|c_i \sim Binomial(c_i, \mathcal{P}_{\ig} d_j ).\]
However, we still need to take into account the fact that $c_i$ is a Poisson-distributed random variable with arrival rate $d_i$. Consequently, the unconditional distribution of $A_{ij}$ is Poisson as well: 
\[ A_{ij} \sim Poisson( d_i d_j \mathcal{P}_{\ig}  ).\]

We prove this fact by multiplying the conditional density of $A_{ij}|c_i$ by the marginal density of $c_i$ to get the joint density of $A_{ij}$ and $c_i$, and then integrating out $c_i$.

\begin{flalign*}
	P(A_{ij},c_i) = \underset{Bin(c_i, d_j P_{\iota\gamma}) }{\underbrace{P(A_{ij}|c_i)}} \quad \times \quad \underset{Poisson(d_i)}{\underbrace{P(c_i)}} \\
\end{flalign*}

Deriving the joint distribution:
\begin{flalign*}
	P(A_{ij},c_i) =& \binom{c_i}{A_{ij}} (d_j P_{\iota\gamma})^{A{ij}}(1-d_j P_{\iota\gamma})^{c_i - A{ij}} \times \frac{d_i^{c_i} \exp{(-d_i})}{c_i!} \\
\end{flalign*}

We want to find out the marginal distribution of $A_{ij}$:
\begin{flalign*}
	P(A_{ij}) &= \sum_{c_i=0}^\infty P(A_{ij},c_i) \\
	&= \sum_{c_i=0}^\infty \binom{c_i}{A_{ij}} (d_j P_{\iota\gamma})^{A{ij}}(1-d_j P_{\iota\gamma})^{c_i - A{ij}} \times \frac{d_i^{c_i} \exp{(-d_i})}{c_i!} \\
	&= \sum_{c_i=0}^\infty \frac{c_i!}{A_{ij}!(di-A_{ij})!} (d_j P_{\iota\gamma})^{A{ij}}(1-d_j P_{\iota\gamma})^{c_i - A{ij}} \times \frac{d_i^{c_i} \exp{(-d_i})}{c_i!} \\
	&= \frac{(d_j P_{\iota\gamma})^{A{ij}}\exp{(-d_i})}{A_{ij}!}  \sum_{c_i=0}^\infty \frac{1}{(di-A_{ij})!} (1-d_j P_{\iota\gamma})^{c_i - A{ij}} d_i^{c_i} \\
\end{flalign*}

If the summation term is equal to 
\begin{equation}\label{eq_bracket}
	\sum_{c_i=0}^\infty \frac{1}{(di-A_{ij})!} (1-d_j P_{\iota\gamma})^{c_i - A{ij}} d_i^{c_i} = d_i^{A_{ij}} \exp{(d_i (1 - d_j P_{\iota\gamma}))}
\end{equation}
then $P(A_{ij}) = \frac{(d_i d_j P_{\iota\gamma})^{A{ij}}\exp{(-d_i d_j P_{\iota\gamma}})}{A_{ij}!}$, i.e. $A_{ij}$ would be Poisson distributed: \\
\[ A_{ij} \sim Poisson(d_i d_j P_{\iota\gamma}) \] 

\vspace{2em}
Proving (\ref{eq_bracket}) is equivalent to proving the following equality:
\begin{flalign*}
	1 =& \frac{1}{d_i^{A_{ij}} \exp{(d_i (1 - d_j P_{\iota\gamma}))}}  \sum_{c_i=0}^\infty \frac{1}{(di-A_{ij})!} (1-d_j P_{\iota\gamma})^{c_i - A{ij}} d_i^{c_i} \\
\end{flalign*}

Proof:
\begin{flalign*}
	& d_i^{-A_{ij}} \exp{(-d_i (1 - d_j P_{\iota\gamma}))}  \sum_{c_i=0}^\infty \frac{1}{(di-A_{ij})!} (1-d_j P_{\iota\gamma})^{c_i - A{ij}} d_i^{c_i} = \\
	&= \sum_{c_i=0}^\infty \frac{\exp{(-d_i (1 - d_j P_{\iota\gamma}))}}{(di-A_{ij})!} (1-d_j P_{\iota\gamma})^{c_i - A{ij}} d_i^{c_i-A_{ij}} \\
	&= \sum_{c_i=0}^\infty \frac{\exp{(-d_i (1 - d_j P_{\iota\gamma}))}}{(di-A_{ij})!} (d_i(1-d_j P_{\iota\gamma}))^{c_i - A{ij}} \\
	&\text{We assume $\lambda = d_i (1 - d_j P_{\iota\gamma})$ for simplicity and we apply a change of variables $z = c_i - A_{ij}$} \\
	&= \sum_{z=0}^\infty \frac{\exp{(-\lambda)}}{z!} \lambda^z \text{, knowing that in our problem $c_i \geq A_{ij}$, i.e. $z \geq 0$}. \\
	&= 1 \\
	&\text{Since we have the p.d.f. of a Poisson r.v. inside the summation, i.e. $z \sim Poisson(\lambda)$  } \square\\ 
\end{flalign*}

Therefore, we have 

\[ A_{ij} \sim Poisson(d_i d_j P_{\iota\gamma}) \qed \]


\clearpage

\section{Soft assignment workers and jobs to worker types and markets}\label{app:network_theory}

In section \ref{sec:model_and_sbm}, at the maximum of our posterior in equation \ref{eq_posterior}, each worker is assigned to only one skill cluster, a process of \textit{hard assignments}. However, it is possible that, given the pattern of worker matches, a particular worker could be revealed to possess certain skills $\i_1$ in most of her matches, and skills $\i_2$ in a few other of her matches. Creating a single worker skill group to accommodate her hybrid skills might not improve model fit if there are only a few workers who exhibit similar matches. Instead, allowing her to have mixed skills $\i_1$ and $\i_2$, i.e. \textit{soft assignment}, with weights according to her matching history, provides further nuanced information to the researcher. In fact, we propose using the Bayesian setup in order to recover these weights. 

It turns out that the posterior $P(\bm{b} | \bm{A}, \bm{g})$ ultimately carries the desired measure of workers' \textit{skill profile} needed to control for workers' unobserved skills in the wage gap estimation. Given a total of $I$ clusters of workers competing for the same jobs in the labor market network, i.e. with similar skills, the posterior distribution provides the chance of each worker to belong to a certain skill cluster, given the worker demographic group $g$ and the entire network $\bm{A}$. More formally, for worker $i$, her \textit{skills profile} is defined as:
\begin{flalign}\label{eq_skills_profile}
    \vec{P}_i := \left[ P(i \in \i_1 | \bm{A}, \bm{g}) \qquad P(i \in \i_2 | \bm{A}, \bm{g}) \qquad \cdots \qquad P(i \in \i_I | \bm{A}, \bm{g}) \right]^T
\end{flalign}

\end{document}